\documentclass[letterpaper,11pt]{article}

\title{Polyhedral Clinching Auctions with a Single Sample}
\author{Ryosuke Sato\thanks{Graduate School of Science and Technology, Keio University, Yokohama, Kanagawa, 223-0061, Japan. ryosuke.sato.517@gmail.com}}
\date{}
\usepackage[top=1truein,bottom=1truein,left=1truein,right=1truein]{geometry}
\usepackage{amsmath}
\usepackage{algorithm}
\usepackage{algorithmic}
\usepackage{comment}
\usepackage{caption}
\usepackage{multirow}
\usepackage{amsthm}
\usepackage{graphicx}
\usepackage{tikz}
\usetikzlibrary{arrows.meta,positioning,calc}
\usepackage[hidelinks]{hyperref}
\usepackage{amsfonts}
\usepackage{natbib}
\setcitestyle{authoryear,open={[},close={]}}
\begin{document}
\theoremstyle{definition}
\newtheorem{definition}{definition}[section]
\newtheorem{proposition}[definition]{Proposition}
\newtheorem{lemma}[definition]{Lemma}
\newtheorem*{main}{Main Theorem}
\newtheorem*{theorem2}{Theorem}
\newtheorem{theorem}[definition]{Theorem}
\newtheorem{corollary}[definition]{Corollary}
\newtheorem{remark}[definition]{Remark}
\newtheorem*{rem}{Remark}
\newtheorem{fact}[definition]{Fact}
\newtheorem{claim}[definition]{Claim}
\newtheorem{observation}[definition]{Observation}
\newtheorem{assmp}[definition]{Assumption}
\newtheorem{example}[definition]{Example}

\maketitle
\begin{abstract}
We address auctions in two-sided markets with budget constraints on buyers, a fundamental setting also crucial for applications such as display advertising. Our goal is to design efficient mechanisms that satisfy dominant strategy incentive compatibility, individual rationality, and budget balance. To overcome the limitations of impossibility theorems, we assume prior knowledge of sellers' valuations and focus on liquid welfare, an efficiency objective that takes budgets into account. Our contributions are twofold: First, we improve the efficiency guarantees of the polyhedral clinching auction by Hirai and Sato (2022). Second, using the reduction method~of D\"{u}tting et al. (2021), we extend the mechanism to an efficient single-sample mechanism for budget-constrained auctions, providing the budget extension of their results. Notably, our results hold even under polymatroid constraints and apply to both divisible and indivisible goods.
\end{abstract}
\section{Introduction}
In recent years, \textit{two-sided markets} have been widely observed in the real world with the spread of platforms such 
as ad exchange platforms (e.g., Xandr), Internet trading platforms (e.g., eBay), and ride-sharing platforms (e.g., Uber).  
Although the theory of \textit{mechanism design} for these markets is a priority for future development, they have problematic structures, i.e., both buyers and sellers have incentives to act strategically to maximize their utility. Mechanisms capable of addressing these strategic actions are necessary for efficient resource allocation in two-sided~markets.

Despite the wealth of knowledge accumulated from auctions in \textit{one-sided markets} (see, e.g., \citep{K2010,NRTV2007}), 
the basic theory for auctions in two-sided markets remains underdeveloped and faces significant challenges. 
A seminal paper \citep{MS1983} revealed that in bilateral trade---the simplest form of a two-sided market---an incentive compatible and individually rational mechanism that maximizes \textit{social welfare} (SW) would run a deficit.

A recent study \citep{DFLLR2021} also revealed that without prior information on valuations, there is no mechanism that satisfies all of the \textit{dominant strategy incentive compatibility} (DSIC), \textit{individual rationality} (IR), \textit{budget balance} (BB), and a constant approximation to the optimal SW. 
Instead, by assuming minimal prior information---known as the \textit{single-sample assumption}---they proposed \textit{single-sample mechanisms} to achieve all of the properties. 
These mechanisms are derived by black-box reductions, which transform any one-sided mechanism into a two-sided one with some loss of efficiency.
They implied two excellent points for the design of efficient mechanisms in two-sided markets in the \textit{absence} of \textit{budget constraints}: 
\begin{itemize}
\item A single sample from each seller's distribution of valuations is sufficient to obtain a constant approximation to the optimal SW while satisfying DSIC, IR, and BB. 
\item Efficient mechanisms in one-sided markets can also be useful in two-sided markets.
\end{itemize}

As these implications hold even in the presence of constraints on the allocation of goods, \textit{budget constraints} would be a natural next challenge. Even in two-sided markets, multi-unit auctions with budgets are important, not only theoretically but also in practice. For instance, in display advertising, advertisers often have maximum possible payments to publishers for a given time period. However, budget constraints have theoretical difficulties, even in one-sided markets. A well-known folklore (see, e.g., \citep{DLN2012, DL2014}) states that even a constant approximation to the optimal SW cannot be achieved by any mechanism that satisfies DSIC and IR. 
Thus, under budget constraints, an alternative efficiency objective is needed.

This study proposes an efficient mechanism for \textit{budget-constrained auctions} in two-sided markets, satisfying all of DSIC, IR, and BB. 
This is the first study to address this issue under the single-sample assumption. 
Building on the first point and excluding budget constraints, this assumption has proven sufficient to design an efficient mechanism.
In addition, following the second point, the candidate for the efficient mechanism  
is the \textit{polyhedral clinching auction}, which~is known to be the most generalized and efficient mechanism for budget-constrained auctions in one-sided markets. 
This mechanism was first proposed by \citet{GMP2015}, enhancing the celebrated~clinching framework of \citet{A2004}.
We show that the above two points are also valid~under budget constraints by extending the polyhedral clinching auction to an efficient single-sample~mechanism.

\subsection{How Do We Measure the Efficiency?}

Evaluating efficiency is a key aspect of this study. In budget-constrained auctions, it is impossible to achieve a constant approximation to the optimal SW while satisfying DSIC and IR. Then, a weaker notion of efficiency, \textit{Pareto optimality} (PO), is often used as an alternative efficiency goal. However, the impossibility theorem \citep{MS1983} still holds even~if we replace optimal SW with PO. Moreover, in the single-sample setting, where average-case analysis is used, approximating PO (e.g., \citep{ILWM2017}) does not intuitively incorporate expectations.

To develop a fundamental theory of auctions in two-sided markets, we need results that is applicable to highly generalized models. These models should allow us to exploit the advantages of polyhedral clinching auctions. This means that certain efficiency guarantees, particularly those designed for symmetric settings (e.g., common budgets or i.i.d. distribution of valuations) as in \citep{DHH2013} and \citep{FH2018}, are not suitable for our asymmetric setting. Thus, we require alternative efficiency targets beyond PO also applicable to such~asymmetries.

\textit{Liquid welfare} (LW) \citep{DL2014,ST2013} is the candidate for our efficiency objective. LW naturally extends SW by incorporating the ability-to-pay (i.e., budgets). While SW is the sum of the willingness-to-pay (i.e., the valuation of the allocated goods), LW is defined as the sum of the admissibility-to-pay, which is the minimum of the ability-to-pay and the willingness-to-pay. It is used as a standard efficiency objective in budget-constrained auctions (e.g.,~\citet{FT2023, FLP2019,LX2017}). 

However, the impossibility theorems \citep{MS1983, DFLLR2021} for SW also hold for LW. Therefore, our best achievable goal is to design \textit{a single-sample mechanism that achieves a constant approximation to the optimal LW while satisfying DSIC, IR, and BB}. This represents a novel and challenging extension of the work by \citet{DFLLR2021} to address budget constraints, emphasizing that our research goals are both fundamental and important.

\subsection{Our Contributions}
We consider budget-constrained auctions with multiple buyers and sellers, where each seller offers multiple units of a homogeneous good. Each seller's valuation is drawn from a distribution, but the auctioneer has access to only a single sample from each seller. This study is the first to address budget-constrained auctions in two-sided markets with limited information. Our main result is:

\begin{main}
\label{DSBBmech/w.samples}
There exists a single-sample mechanism that satisfies all of DSIC, IR, and BB.
Moreover, the mechanism achieves LW more than 1/4 of the optimal LW, and achieves SW more than 1/2 of the optimal LW, in expectation.  
\end{main}
This theorem shows that we have successfully achieved our goal of a constant approximation to the optimal LW while also providing an SW guarantee.
The use of the optimal LW as a benchmark for SW was first proposed by \citet{ST2013}, and this study is the first to apply this benchmark to clinching auctions. Since the optimal LW can be interpreted as the optimal SW when valuations are modified to budget-additive valuations, this comparison provides valuable insights.

Since the existing upper bounds for both guarantees are 1/2 from the results of \citet{DFLLR2021} and \citet{LRW2023}, our SW guarantee even reaches the best possible. Furthermore, in the case of bilateral trade, we also show that the LW of our mechanism achieves this ratio. This confirms the efficiency of our mechanism and highlights its effectiveness.

A notable feature of this result lies in its \textit{generality}. Despite the challenges inherent in our setting, we successfully achieve our goal under both \textit{divisible} and \textit{indivisible} cases with \textit{polymatroid} constraints on the allocation of goods. In fact, our model captures a two-sided extension of many existing works on budget-constrained auctions; e.g., multi-unit auctions~\citep{DLN2012}, matching markets \citep{FLSS2011, BHLS2015}, and adwords auctions \citep{GMP2015}. Our model is also a budget extension of reservation exchange markets~\citep{GLMNP2016}.

Our single-sample mechanism builds on the reduction method of \citet{DFLLR2021}, which extends the VCG mechanism (for non-strategic sellers) to single-sample settings. Given that the polyhedral clinching auction yields the VCG outcome in the absence of budgets, it seems reasonable to use this mechanism. However, the existing efficiency guarantee for the mechanism is limited to PO, which does not fit our purpose.
To lay the groundwork for the efficiency analysis of our mechanism, we establish \textit{tight} LW and SW guarantees for the polyhedral clinching auction:
\begin{theorem2}[Theorems \ref{LWofHS} and \ref{SW}, Informal]
\label{liquid-welfare/w.o.samples}
The polyhedral clinching auction (for divisible goods)
achieves an LW of more than 1/2 of the optimal LW, and an SW of more than the optimal LW.
\end{theorem2}

Our LW guarantee drastically extends the result of \citet{DL2014}, which focuses on multi-unit auctions in one-sided markets, while preserving the approximation ratio. This is the first LW guarantee under polymatroid constraints. Our technical contribution lies in new invariant properties for analyzing the remaining goods and payments of the mechanism, which have not been explored in previous works, even in special cases of our setting. Then, by showing that D\"{u}tting's reduction method also applies to LW guarantees, we derive our main theorem from this theorem.

\subsection{Related Works.}

\paragraph{Auctions in Two-sided Markets (without Budgets).}
The well-known impossibility theorem \citep{MS1983} set the research direction towards the design of approximately efficient mechanisms that satisfy other desirable properties. In bilateral trade, recent efforts have focused on improving the approximation ratio of SW (e.g., \citep{BD2021,KPV2022,LRW2023,DS2024}) or gains-from-trade (e.g., \citep{BCWZ2017,DMSW2022,F2022}), as summarized in \citet{LRW2023}. More general settings with constraints on the allocation of goods \citep{BKLT2016, BK2021} and combinatorial structures \citep{BGKLR2020} have gained attention due to their growing importance. These studies consider Bayesian settings, where the auctioneer can access to the distributions of valuations. In prior-free settings, asymptotically efficient mechanisms satisfying DSIC, IR, and BB have been proposed for double auctions (e.g., \citep{MA1992}) and multi-unit auctions (e.g., \citep{SHA20181,SHA20182}). However, \citet{DFLLR2021} showed the impossibility of achieving a constant approximation to the optimal SW while satisfying DSIC, IR, and BB. They then proposed efficient single-sample mechanisms that use black-box reduction methods (as in \citet{DRT2017}).
Single-sample settings are often considered when assuming the minimum prior information in auctions (e.g., \citep{DRY2015,KPV2022}) as well as in prophet inequalities (e.g., \citep{AKW2014, AKW2019, CDFS2019, CCES2020}). Recently, \citet{CW2023} and \citet{BFN2024} also addressed the settings with more than~one~sample.

\paragraph{Clinching Auctions with Budgets.}
\citet{DLN2012} showed that a variant of clinching auctions \citep{A2004} is one of the candidates for efficient mechanisms in budget-constrained auctions.
Subsequent studies (e.g., \citep{BCMX2010,BHLS2015,DHS2015,FLSS2011,GMP2014, GMP2015,GMP2020,HS2022,HS2023}) 
have generalized the model and mechanism, 
which are the cornerstones of our single-sample mechanism.
After the first version of this paper was uploaded to arXiv,
\citet{HS2023} proposed the polyhedral clinching auction for indivisible goods and 
used our techniques for their LW and SW guarantees.
Their result is further extended to the single-sample setting in this study.

\paragraph{Another Approaches in Budget-constrained Auctions.}
A number of studies (e.g., \citep{A2006,CGMW2021,FHL2023,V2020}) have also investigated the effectiveness of simple mechanisms in budget-constrained auctions, providing the guarantees on welfare and revenue.
Also, \citet{WLL2018} focused on procurement mechanisms in two-sided markets.
Their setting is quite different from ours in that the payment of each buyer is independent of her utility if it is within the budget, and that she does not report the bid corresponding to her valuation.

\subsection{Organization of the paper.}
Section 2 presents our model.
Section 3 introduces the polyhedral clinching auction and its properties.
Section 4 provides efficiency guarantees for the mechanism.
Section 5 extends the mechanism to single-sample settings using D\"{u}tting's reduction method and presents the main result.
Section~6 explores further extensions and applications of our results.
Section 7 examines whether our efficiency guarantees achieve the best possible and suggests directions for future research.
Section~8 proves a key result for the LW guarantee in Section 4.
Some proofs are given in the~appendix.

\rem{Our main theorem holds even in the case where the allocation of goods is constrained by polymatroids, or the goods are indivisible, or both. For the sake of simplicity, we limit the arguments in Sections 2 through 5 to the divisible case without polymatroid constraints. Then, in Section 6, we explain how we can extend our results to these broad settings.}

\paragraph{Notation.}
Let $\mathbf R_+$ and $\mathbf R_{++}$ 
denote the set of nonnegative and positive real numbers, respectively, 
and let $\mathbf Z_+$ denote the set of nonnegative integers.
For a set $E$, let $\mathbf R^E_+$ denote the set of all functions from $E$ to $\mathbf R_+$.
For $w\in \mathbf R^E_+$, we often denote $w(e)$ by $w_e$, and write as $w= (w_e)_{e\in E}$. 
For $F\subseteq E$, let $w(F)$ denote the sum of $w(e)$ over $e\in F$ and 
$w|_{F}$ denote the restriction of $w$ to~$F$.
We denote a singleton $\{e\}$ by $e$. 
A bipartite graph consisting of two disjoint sets $N, M$ and an edge set $E \subseteq N \times M$ is denoted by $(N,M,E)$. An edge $(i,j) \in E$ is denoted by $ij$. For a node $i \in N\cup M$, let $E_i\subseteq E$ denote the set of edges connected to $i$. Similarly, let $E_S$ denote the set of edges connected to nodes in $S\subseteq N\cup M$. For $S\subseteq N$, let $M_S$ denote the set of nodes in $M$ adjacent to~$S$.

\section{Our Model}
	Consider a market consisting of $n$ buyers and $m$ sellers, where $n$ and $m$ are positive integers. 
	The market is represented by a bipartite graph $(N,M,E)$, where the set of nodes $N:=\{1,2,\ldots,n\}$ (resp. $M:=\{1,2,\ldots,m\}$) denotes a set of buyers (resp. sellers), and
	buyer $i\in N$ and seller $j\in M$ are adjacent if and only if they intend to transact with one~another.

	We consider a multi-unit auction of homogeneous goods.
	Each buyer $i$ and seller $j$ has a valuation, 
	$v_i \in \mathbf R_{++}$ and $\rho_j \in \mathbf R_{++}$ respectively, for a unit. 
	They report a bid, $v'_i \in \mathbf R_{++}$ for buyers and $\rho'_j \in \mathbf R_{++}$ for sellers, to the auctioneer. 
	Each participant (buyer or seller) acts strategically to maximize her utility.
	Moreover, buyer $i$ has a budget $B_i \in \mathbf R_+$ for the maximum possible payment and 
	seller $j$ has $s_j\in \mathbf R_{+}$ units of the goods.
	An allocation $\mathcal A$ is a triple $\mathcal A:=(w, p, r)$ of a transaction vector $w=(w_{ij})_{ij\in E}$, 
	a payment vector $p=(p_{i})_{i\in N}$, and a revenue vector $r=(r_{j})_{j\in M}$, 
	where $w_{ij}\in \mathbf R_{+}$ denotes the amount of goods transacted between buyer $i$ and seller $j$, 
	$p_i\in \mathbf R_{+}$ denotes the payment of buyer $i$, and 
	$r_j\in \mathbf R_{+}$ denotes the revenue of seller~$j$.
	The allocation must satisfy the budget constraint $p_{i}\leq B_i\ (i\in N)$ 
	and the supply constraint $w(E_j)\leq s_j\ (j\in M)$.

	The utilities of the participants are determined by the allocation $\mathcal A$.
	For a transaction vector $w$, we define a vector 
	$x:=(x_i)_{i\in N}\in \mathbf R^{N}_{+}$ by $x_i:=w(E_i)$ for each $i\in N$, 
	which represents the number of goods allocated to buyer $i$.
	Then, for allocation $\mathcal A$, the utility $u_i(\mathcal A)$ of buyer $i$ is defined by 
	\[
	u_i(\mathcal A):=
	\begin{cases}
	\displaystyle 
	v_i x_i-p_i\quad\,  {\rm if}\ p_i\leq B_i, \\
	-\infty \qquad\quad  {\rm otherwise}, 
	\end{cases}
	\]
	which is the valuation for the allocated goods minus her payment if the payment is within the budget and, 
	otherwise, the utility goes to -infinity. 
	The utility of seller $j$ is defined by $u_j(\mathcal A):=r_j-\rho_j w(E_j)$, 
	which is her revenue minus the total valuation of her goods sold in the auction.\footnote{We modify the utility of the seller from \citet{HS2022} so that the utility of not participating becomes~zero.}
	
	The information $\mathcal I$ to which the auctioneer has access is defined by 
	$\mathcal I:=((N,M,E)$, $\{v'_i\}_{i\in N}$, $\{B_i\}_{i\in N}$, $\{\rho'_j\}_{j\in M}$, $\{s_j\}_{j\in M}$). Due to the impossibility theorems by \citet{DLN2012} and \citet{FLSS2011}, we assume that the budgets $\{B_i\}_{i\in N}$ and the edge set $E$ are \textit{public} information.
	The valuation $v_i$ (resp. $\rho_j$) of buyer $i$ (resp. seller $j$) 
	is the private knowledge that only the participant has access to. 
	Then, the triple of ($\mathcal I$, $\{v_i\}_{i\in N}$, $\{\rho_j\}_{j\in M}$) can be regarded as the input to this model.
	A mechanism $\mathcal M$ is a map that outputs an allocation $\mathcal A$ 
	from information $\mathcal I$.
	Then, we consider the following properties for the mechanism:
	\begin{itemize}
	\item Dominant strategy incentive compatibility\,(DSIC):\ \ 
	It is the best strategy for each participant to report her true valuation.
	For every input, it holds 
	$\displaystyle u_i(\mathcal M(\mathcal I))\leq u_i(\mathcal M(\mathcal I_i))\  (i\in N\cup M)$,
	where $\mathcal I_i$ is obtained from $\mathcal I$ 
	by replacing the bid of participant $i$ with her valuation.
	\item Individual rationality\,(IR):\ \ 
	For each $i\in N\cup M$, there is a bid such that $i$ 
	receives non-negative utility.
	If DSIC holds, IR is expressed as 
	$\displaystyle u_i (\mathcal M(\mathcal I_i))\geq 0 \ (i\in N\cup M)$ for every~input.
	\item Weak budget balance\ (WBB): 
	The total payments exceed the total revenue, i.e.,~$p(N)\geq r(M)$.
	\item Strong budget balance\ (SBB): 
	The total payments equal the total revenue, i.e.,~$p(N)= r(M)$.
	\end{itemize}

	Efficiency is a key concept of our study, and the standard measure is  
	social welfare (SW), which is defined as the sum of the valuations of the allocated goods for each participant, i.e.,
	\[
	{\rm SW}(\mathcal A):= \sum_{i\in N}v_i x_i+\sum_{j\in M}\rho_j (s_j-w(E_j))
	\]
	for an allocation $\mathcal A=(w,p,r)$.\footnote{The value $\sum_{i\in N}v_i x_i-\sum_{j\in M}\rho_j w(E_j)$ is also called \textit{gains-from-trade}, a commonly used efficiency objective. In general, providing theoretical guarantees for gains-from-trade is more challenging than for SW.}
	However, it is well known that in the presence of budgets, 
	no mechanism can achieve a constant approximation to the optimal SW 
	accompanied by DSIC and IR (see Lemma 2.2 in \citet{DL2014}).
	This impossibility theorem also holds in our setting, and thus an alternative goal of efficiency is needed.
	Liquid welfare (LW) is a natural extension of SW and is a standard efficiency objective in budget-constrained auctions. LW is defined by 
	\[
	{\rm LW}(\mathcal A):= \sum_{i\in N}\min(v_i x_i, B_i)+\sum_{j\in M}\rho_j (s_j-w(E_j)),
	\]
	which represents the total admissibility-to-pay. 
	In one-sided markets, some mechanisms achieve a constant approximation to the optimal LW with DSIC and IR (e.g., \citep{DL2014}). 	
	However, a constant approximation to the optimal LW accompanied by DSIC, IR, and BB is impossible in two-sided markets without priors due to Theorem 1 of \citet{DFLLR2021}.\footnote{Their impossibility theorem applies to indivisible goods, and since their proof for randomized mechanisms also serves as a proof for divisible goods, it covers both divisible and indivisible goods.}
	
	To circumvent the above limitations, we assume specific prior information on sellers' valuations:
	\begin{itemize}
	\setlength{\itemindent}{-5pt}
	\item In Sections 3 and 4, sellers are assumed to report their valuations truthfully. 
	\item In Section 5, sellers' valuations are drawn independently from probability distributions and the auctioneer has access to only a single sample from each seller's distribution.
	\end{itemize}
	
	\rem{Pareto optimality (PO) has also been used as an efficiency objective 
	in budget-constrained auctions. However, the impossibility theorem by \citet{MS1983} 
	implies that PO cannot be achieved with DSIC, IR, and BB 
	in our setting of Section 5.}

\section{Polyhedral Clinching Auctions with Truthful Sellers}
	In the reduction method of \citet{DFLLR2021} used in Section 5, we need a mechanism that is applicable to the market with multiple non-strategic sellers with different bids. This motivates us to focus on the polyhedral clinching auction by \cite{HS2022}, which extends the one by \citet{GMP2015} to such setting.
	In this section, we introduce the polyhedral clinching auction by \citet{HS2022} and provide some basic 
	properties. Throughout Sections 3 and 4, sellers are assumed to truthfully report their bids, 
	i.e., $\rho'_j=\rho_j$ for each $j\in M$. 
\subsection{Preprocessing}
	Throughout this paper, the following preprocessing is often used, where $m$ \textit{virtual buyers} are added and then the market is modified.
	Each virtual buyer, associated with seller $j$, is labeled buyer $n+j$ 
	and is only adjacent to seller $j$. 
	Then, the set of buyers $N$ is modified to $N:=\{1,2,\ldots,n+m\}$, 
	and the set of edges $E$ is modified to $E:=\bigcup_{j\in M} E_j\cup \{(n+j)j\}$.
	We assume that $v'_{n+j}=v_{n+j}=\rho_j$ and $B_{n+j}=\infty$ for each $j\in M$.
	This modification shifts sellers' bids/valuations toward virtual buyers.
	Then, we consider this modified market and do not use sellers' bids/valuations.
	Note that these modifications are also used in \citet{GLMNP2016} and \citet{HS2022}.
	
	Suppose that all goods are sold to buyers at the end of the auction.
	Then, for seller $j\in M$, the goods allocated to virtual buyer $n+j$  
	are considered as the unsold goods of $j$.
	The true utility of seller $j$ can be calculated as the sum of the utilities of seller $j$ and the virtual buyer $n+j$ by
	\[
	\left(r_j-\rho_j w(E_j)\right)+\left(v_{n+j} x_{n+j}-p_{n+j}\right)=(r_j-p_{n+j})-\rho_{j} \left(w(E_j)-x_{n+j}\right), 
	\]
	where the first term on the right-hand side is the revenue 
	from non-virtual buyers, and the second term is the valuation of the goods sold to non-virtual buyers.
	
	In this market, SW is modified to ${\rm SW}(\mathcal A):= \sum_{i\in N}v_i x_i$. 
	Since the budgets of virtual buyers are infinite, 
	LW is also modified to ${\rm LW}(\mathcal A):= \sum_{i\in N}\min(v_i x_i,B_i)$. 
	These are the ones defined for one-sided markets, and thus virtual buyers make our two-sided market closer to a one-sided~market.

\subsection{Mechanism}
	The full description of the mechanism is provided in Algorithm \ref{HS_PCA}. 
	We denote the mechanism by ${\rm PCA}((N,M,E),v',\rho', B, s)$, where $(N,M,E)$ denotes the (original) market, 
	$v':=\{v'_i\}_{i\in N}$ denotes the set of buyers' bids, $\rho':=\{\rho'_j\}_{j\in M}$ denotes the set of sellers' bids, 
	$B:=\{B_i\}_{i\in N}$ denotes the set of buyers' budgets, 
	and $s:=\{s_j\}_{j\in M}$ denotes the set of the number of goods supplied by sellers. 
	\begin{algorithm}[htb]
	\caption{Polyhedral\ clinching\ auction by \citet{HS2022}: ${\rm PCA}((N,M,E),v', \rho', B, s)$}
	\label{HS_PCA}                           
	\begin{algorithmic}[1]
	 \STATE Add virtual buyers $ V$ and construct a modified market $(N, M, E)$. 
	  \STATE $c_i:=0,\ d_i:=\infty\ 
	  (i\in N)\ {\rm and}\ l:=1$. $w_{ij}:=0\ (ij\in E)$,\ $p_i:=0,\ (i\in N)$,\ and $r_j:=0\ (j\in M)$.
	  \WHILE{$d_i\neq 0$ for some $i\in N$}
	\STATE Clinching$(s, (w, p, r), d, c)$.
	\STATE $c_{l}:=c_{l}+\varepsilon$.
	  \STATE $ d_l:=
	 \begin{cases}
	(B_l-p_l)/c_l\quad {\rm if}\  c_l<v'_l,\\
	0 \quad\qquad\qquad\ \  {\rm otherwise}.
	\end{cases}$
	   \STATE $l:=l+1\ {\rm mod\ } m+n$.
	 \ENDWHILE
	\STATE  $w^{\rm f}_{ij}:=w_{ij}\ (ij\in E)$,\ $p^{\rm f}_i:=p_i\ (i\in N)$,\ {\rm and} \  $r^{\rm f}_j:=r_j\ (j\in M)$.
	\STATE  Output the final allocation after cancelling all transactions between virtual buyers and~sellers.
	\end{algorithmic}
	\end{algorithm}
	
The variables and the parameter are explained as follows:
\begin{itemize}
\item $(w, p, r)$ is the current allocation, 
where $w_{ij}\in \mathbf R_{+}$ is the number of goods transacted between buyer $i$ and seller $j$, 
$p_{i}\in \mathbf R_{+}$ is the payment of buyer $i$, and $r_{j}\in \mathbf R_{+}$ is the revenue of seller~$j$.
\item $c:=(c_i)_{i\in N}\in \mathbf R^N_{+}$ is the transaction price per unit of buyers. For each $i$, the price $c_i$ starts at zero and increases by $\varepsilon$ at each step. We assume that all values in $v'$ are multiples of $\varepsilon$.
\item $d:=(d_i)_{i\in N}\in \mathbf R^N_{+}$ is the demands of buyers.  For buyer $i$, the value $d_i$ represents the number of maximum  amounts that $i$ can clinch at the current price. Specifically, buyers with zero demand will not clinch any good, and we call a buyer \textit{dropping} if her demand reaches~zero. 
\item $l$ is the buyer whose price is increased in line 5 of the current iteration.
\end{itemize}
	
	In line 1, the preprocessing in Section 3.1 is performed to construct a modified market $(N, M, E)$, where 
	the set $V$ denotes the set of virtual buyers. 
	Then, the following ascending auction is conducted in this new market. 
	Initially, each buyer $i$ has a price clock $c_i$ that starts at $0$ and gradually increases by $\varepsilon$, and the allocation $(w,p,r)$ is all equal to $0$.
	Each iteration comprises the following steps: 
	In line 4, if a certain condition (described below) is satisfied, buyers clinch some amount of goods at the current price and the allocation $(w,p,r)$ is updated according to the transactions. In line~5, the price clock $c_l$ of buyer $l$ is incremented by $\varepsilon$, where $l$ is determined in order. In line 6, her demand $d_l$ is also updated. If there are buyers with positive demands, the auction proceeds to the next iteration.
	Otherwise, in line 10, Algorithm~\ref{HS_PCA} terminates and outputs the final allocation 
	after canceling all virtual transactions.
	\begin{rem}
	Throughout this paper, we use $(w^{\rm f}, p^{\rm f}, r^{\rm f})$ in line 9 as the final allocation in the analysis, rather than the output in line 10. By the preprocessing in Section 3.1, the cancellation in line 10 does not affect SW or LW, and thus $(w^{\rm f}, p^{\rm f}, r^{\rm f})$ accurately reflects the efficiency of the mechanism.
	\end{rem}

	The description of the clinching procedure (line 5 of Algorithm \ref{HS_PCA}) is provided in Algorithm~\ref{HS_clinch}.
	Intuitively, buyers clinch the maximum~amount of goods that does not affect other buyers.
	\begin{algorithm}[htb]
	\caption{Clinching$(s, (w, p, r), d, c)$}
	\label{HS_clinch}                           
	\begin{algorithmic}[1]
	  \FOR{$i=1,2,\ldots,n+m$} 
	  \STATE Clinch a maximal increase 
	  $\xi_i:=(\xi_{ij})_{ij\in E_i}$ satisfying $P^i_{w,d}(\xi_{i})=P^i_{w,d}(0)$ (see below).
	  \STATE $\displaystyle p_i:=p_i+c_i \xi_i(E_i)\ (i\in N).$
	  \STATE $w_{ij}:=w_{ij}+\xi_{ij}\ \  (ij\in E_i)$,\ \  $ d_i:=
	 \begin{cases}
	(B_i-p_i)/c_i\quad {\rm if}\  c_i<v'_i \\
	0 \quad\qquad\qquad\ \  {\rm otherwise}
	\end{cases}$\ $(i\in N).$
	  \ENDFOR
	  \STATE $\displaystyle r_j:=r_j+\sum_{ij\in E_j} c_i \xi_{ij}\quad (j\in M).$
	\end{algorithmic}
	\end{algorithm}
	
\begin{figure}[htbp]
	\centering
	\includegraphics[width=130mm]{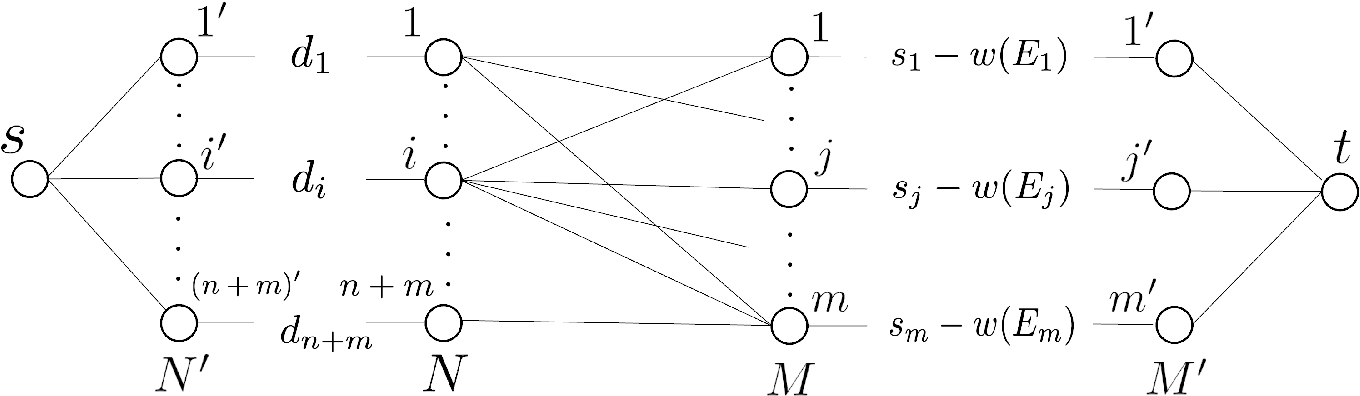}
	\caption{The resulting network $\mathcal N$.}
	\label{network}
	\end{figure}

	To describe the method, we construct the following network:
	Suppose that each edge in $E$ is directed from $N$ to $M$. 
	For each $i\in N$, consider its copy $i'$ and connect 
	it with a directed edge $i'i$. 
	Add a source node $s$ and connect it with 
	each $i'$ through a directed edge $si'$. 
	For each $j\in M$, consider its copy $j'$ and connect it with a directed edge $jj'$.  
	Also, add a sink node $t$ and connect each $j'$ 
	to it with a directed edge $j't$. 
	The edge capacities are defined~by 
	\begin{align*}
	{\rm cap}(e):=
	\begin{cases}
	d_i&{\rm if}\ e=i'i\ {\rm for}\ i\in N,\\
	s_j-w(E_j)&{\rm if}\ e=jj'\ {\rm for}\ j\in M,\\
	\infty&{\rm otherwise}.
	\end{cases}
	\end{align*}
	The resulting network $\mathcal N$ is illustrated in Figure \ref{network}.
	
	For a vector $\kappa_i:=(\kappa_{ij})_{ij\in E_i}\in \mathbf R^{E_i}_+$, 
	we define a polytope $P^i_{w,d}(\kappa_{i})$ by 
	\[
	P^i_{w,d}(\kappa_{i}):=\{u\in \mathbf R^{N\setminus i}_+\ |\ \exists \psi:{\rm flow\ in\ network}\ \mathcal N, 
	\psi(i'i) = \kappa_i(E_i), \psi(k'k) = u_k\ (k\in N\setminus i)\},
	\]
	which represents the possible future transactions of buyers 
	$N\setminus i$ if buyer $i$ clinches $\kappa_i\in \mathbf R^{E_i}_{+}$ in this iteration. 
	Then, the clinching condition is described as 
	$P^i_{w,d}(\kappa_{i})=P^i_{w,d}(0)$.
	The set of $\kappa_i$ that satisfies the condition is known as the \textit{clinching polytope}, i.e., 
	$P^i_{w,d}:=\{\kappa_i\in \mathbf R_{+}^{E_i} \colon P^i_{w,d}(\kappa_{i})=P^i_{w,d}(0)\}$.
	In line 2 of Algorithm \ref{HS_clinch}, 
	a maximal vector $\xi_i=(\xi_{ij})_{ij\in E_i}$ is chosen from $P^i_{w,d}$.

	Let $f:2^N\to\mathbf R_+$ be a monotone submodular function defined by 
	$f(S):=\sum_{j\in M_S}s_j\ (S\subseteq N)$.\footnote{A monotone submodular function $f:2^N\to\mathbf R_+$ is a function that satisfies (i) $f(\emptyset)=0$, (ii) $f(S)\leq f(T)$\ ($S\subseteq T\subseteq N$), and (iii) $f(S\cup i)-f(S)\geq f(T\cup i)-f(T)\ (S,T\subseteq N, S\subseteq T, i\in N\setminus T)$.}
	This function represents the constraint on the number of goods allocated to buyers, induced by the supply constraint.\footnote{For a feasible allocation $w$, the vector $x$ with $x_i=w(E_i)\ (i\in N)$ satisfies $x(S)\leq f(S)$\ $(S\subseteq N)$. Also, if this inequality holds for some $x$, there exists an allocation $w$ with $x_i=w(E_i)\ (i\in N)$ that satisfies the supply constraint.}
	The key features of Algorithm \ref{HS_clinch} are then summarized as follows:

	\begin{theorem}[\citet{HS2022}]
	\label{clinch_sec3}
	In Algorithm \ref{HS_clinch}, the following holds: 
	\begin{itemize}
	\item[(i)] A vector $\xi_i\in P^i_{w,d}$ can be computed in polynomial time.
	\item[(ii)] For any $\xi_i\in P^i_{w,d}$, 
	it holds $\xi(E_i)=f_{x,d}(N)-f_{x,d}(N\setminus i)\ (i\in N)$, where $f_{x,d}$ is also a monotone submodular function 
	and represents the number of goods that buyers $S$ can transact in the future (under $x$ and~$d$), expressed as
	\begin{equation}
	\label{remnant}
	f_{x,d}(S):=\min_{S'\subseteq S}\{\min_{S''\supseteq S'}\{f(S'')-x(S'')\}+d(S\setminus S')\}\quad (S\subseteq N).
	\end{equation}
	\item[(iii)] For each seller $j\in M$, when $c_{n+j}<\rho_j$, then $\xi_{ij} = 0$ for all $ij \in E_j\setminus (n + j)j$.
	\end{itemize}
	\end{theorem}
	\begin{rem}
	Theorem \ref{clinch_sec3} (i) shows that each iteration can be computed in polynomial time. While there is no non-trivial bound on the number of iterations in Algorithm~\ref{HS_PCA}, in the indivisible case (explained in Section 6.2), the number of iterations is bounded by a polynomial in the number of participants and goods. In this case, the entire procedure runs in polynomial time.
	\end{rem}

	We now summarize the key properties of Algorithm~\ref{HS_PCA}. Let $x^{\rm f} \in \mathbf{R}^N_+$ denote the final allocation of goods for buyers (including virtual buyers), where $x^{\rm f}_i = w^{\rm f}(E_i) \ (i \in N)$. In the following arguments, by DSIC for buyers, we can assume without loss of generality that $v'_i = v_i$ for each $i \in N$.
	\begin{theorem}[\citet{HS2022}]
	\label{properties_sec3}
	In Algorithm \ref{HS_PCA}, the following holds: 
	\begin{itemize}
	\item[(i)] The mechanism satisfies all of DSIC for buyers, IR, and~SBB. 
	\item[(ii)] All goods are sold at the end of the auction, i.e., $x^{\rm f}(N)=f(N)=\sum_{j\in M}s_j$. 
	\end{itemize}
	\end{theorem}	

Next, we introduce a structural property that provides further insight into the final allocation. Specifically, it implies that (i) buyers dropping at higher prices are allocated more goods, and (ii) buyers whose valuations are higher than the price at their dropping exhaust their entire budgets. This property plays a crucial role in efficiency arguments for clinching auctions, particularly in establishing PO. In the rest of this paper, we refer to a set $\mathcal T\subseteq N$ as \textit{tight} if $x^{\rm f}(\mathcal T)=f(\mathcal T)$.

\begin{figure}[htbp]
	\begin{center}
	\includegraphics[width=110mm]{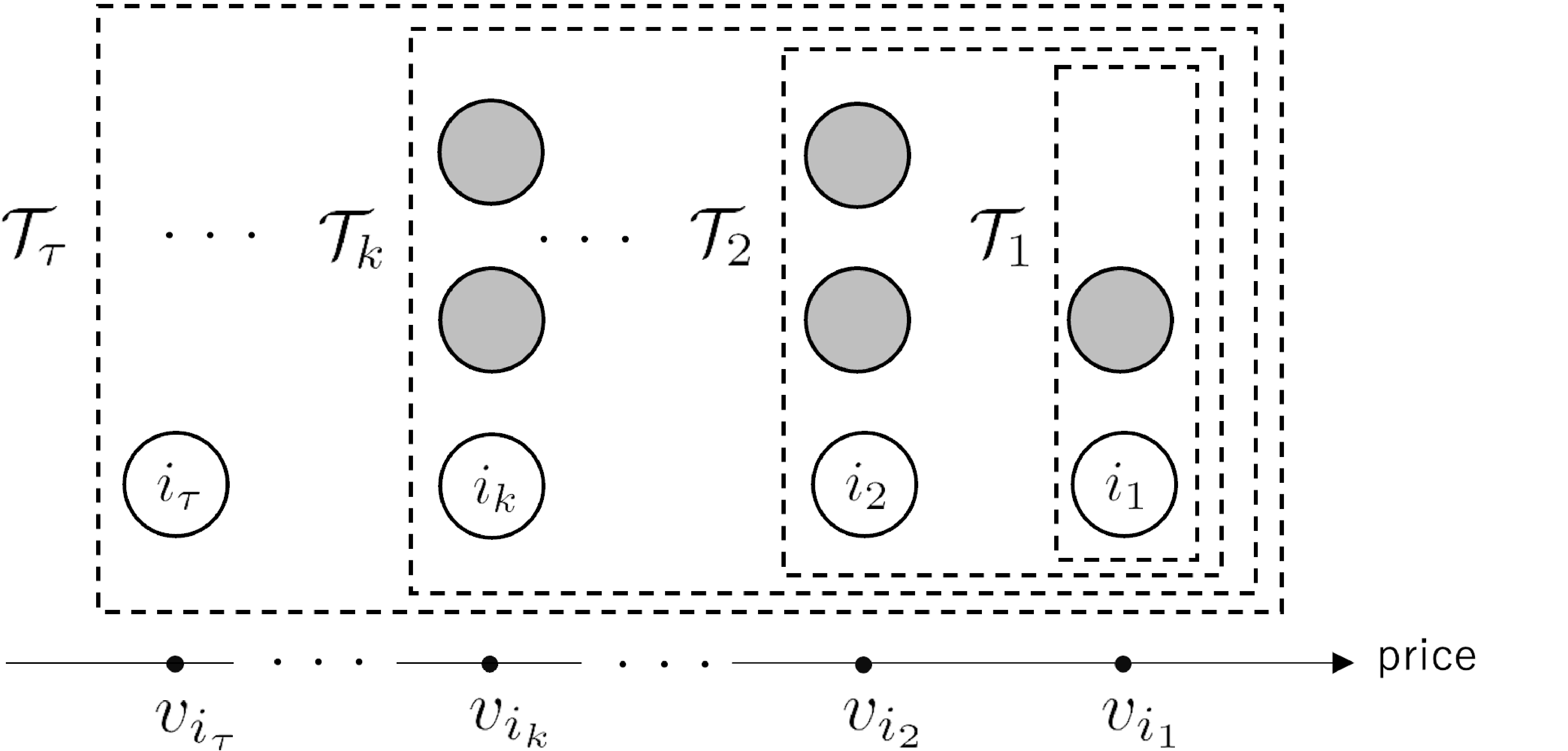}
	\caption{Illustration of the chain of tight sets and the dropping of buyers in Algorithm \ref{HS_PCA}.
	The white circles represent buyers in $\{i_1, i_2, \ldots, i_\tau\}$ and the gray shaded circles represent other buyers.}
	\label{tightsets}
	\end{center}
	\end{figure}

	\begin{proposition}[\citet{GMP2014,GMP2015, HS2022}]
	\label{dropping} 
	Let $i_1,i_2,\ldots,i_\tau$ be the buyers dropped in line 6 of Algorithm \ref{HS_PCA}, sorted in reverse order of  	
	dropping, i.e., $v_{i_1}\geq v_{i_2}\geq\ldots\geq v_{i_\tau}$. For each $k\in\{1,2,\ldots,\tau\}$, 
	let $\mathcal T_k$ denote the set of 
	buyers with positive demands just before the dropping of buyer $i_k$. Then, the following holds: 
	 \begin{enumerate}
	 \item[(i)] $\emptyset= \mathcal T_0\subset \mathcal T_1\subset  \mathcal T_2\subset\cdots\subset  \mathcal T_\tau=N$ is a chain of tight sets, i.e., $x^{\rm f}(\mathcal T_k)=f(\mathcal T_k)$ for each~$k$.

	 \item[(ii)] For each $k$, it holds that $i_k\in \mathcal T_k\setminus \mathcal T_{k-1}$ and $p^{\rm f}_{i_k}<B_{i_k}$.
	 \item[(iii)] For each $k$ and $i\in \mathcal T_k\setminus (\mathcal T_{k-1}\cup i_k)$, 
	 it holds $v_i\geq v_{i_k}$ and $p^{\rm f}_i=B_i$. In particular, buyer $i$ drops out in line 4 of the iteration just after the dropping of buyer $i_k$ in line 6.
	 \end{enumerate}
	\end{proposition}
Figure \ref{tightsets} visualizes the chain of tight sets. 
For each $k$, the set $\mathcal{T}_k$ contains buyers with positive demands just before 
$i_k$ drops out in line 6 of Algorithm \ref{HS_PCA}.
Buyer $i_k$ drops out with some budget remaining, i.e., $p^{\rm f}_{i_k} < B_{i_k}$, while each buyer  
$i \in \mathcal{T}_k \setminus (\mathcal{T}_{k-1} \cup i_k)$ exhausts her budget, i.e., $p^{\rm f}_i = B_i$. 
	
\section{Efficiency Guarantees for Polyhedral Clinching Auctions}
	One issue with our single-sample extension is that the existing efficiency guarantees for the polyhedral clinching auction rely on PO, which cannot be integrated with the reduction method of \citet{DFLLR2021}. In this section, we establish two types of efficiency guarantees for the mechanism.

\subsection{Optimal Liquid Welfare}
	The optimal LW value is used as a benchmark for our efficiency guarantees.
	It can be interpreted as the optimal SW value 
	if the valuation function of each buyer is modified to a budget additive function. 
	Note that this modification has been considered in previous studies 
	(e.g., \citep{LLN2006}) to reduce budget-constrained auctions to auctions~without budgets. 
	
	For preparation of the efficiency guarantees, we construct 
	an allocation that achieves the optimal LW value. Since LW depends only on the total number of goods allocated to buyers, we focus on this quantity and we call it the {\it LW optimal allocation}.
	Given that the auctioneer has access to participants' valuations, 
	this allocation can be constructed as follows:
	First, the preprocessing in Section 3.1 is performed to construct a 
	modified market $(N, M, E)$ with virtual buyers.
	The goods are then allocated to buyers based on the descending order of their valuations.
	For each $i\in N$, the minimum of $B_i/v_i$ and maximum possible amount  
	in terms of the supply constraint is allocated. 
	For buyer $i$, define a set $H_i\subseteq N$ by 
	$H_i:=\{k\in N \colon v_k>v_i\} \cup \{k\in N \colon v_k=v_i\ {\rm and}\ k>i\}$.
	Then, an explicit form of the LW optimal allocation is expressed by the following.
	Note that the proof uses some basic facts on polymatroid theory and thus is given in Appendix C.
	\begin{proposition}
	\label{optimal}
	An LW optimal allocation $x^*:=(x^*_i)_{i\in N}$ is recursively given by 
	\[
	 x^{*}_i=\min\left(B_i/v_i,{\min_{H\subseteq H_{i}}\{f(H\cup i)-x^{*}(H)}\}\right)\ (i\in N).
	\]
	In this allocation, it holds $x^*(S)\leq f(S)\ (S\subseteq N)$ and $x^*(N)=f(N)\, (=\sum_{j\in M}s_j)$.
	Moreover, for non-virtual buyer $i$ with $v_i\leq\min_{j\in M}\rho_j$, it holds $x^{*}_i=0$.
	\end{proposition}
	
	Using this LW optimal allocation, we illustrate the relationship between $x^{*}$ and $x^{\rm f}$. 
	Recall that $x^{\rm f}$ denotes the number of goods allocated to buyers 
	in line 9 of Algorithm \ref{HS_PCA}, i.e., $x^{\rm f}_i=w^{\rm f}(E_i)\ (i\in N)$.
	The following proposition shows that buyers allocated more goods in the mechanism than in the LW optimal allocation are also allocated their full capacity in the LW optimal allocation.

	\begin{proposition}
	\label{x>xopt}
	For buyer $i\in N$, if $x^{\rm f}_i>x^*_i$, then it holds $x_i^*=B_i/v_i.$
	Moreover, for any virtual buyer $i\in  V$, it holds $x^{\rm f}_i\leq x^*_i<B_i/v_i$.
	\end{proposition}
	
	\begin{proof}
	Suppose that there exists a buyer $\ell$ with $x^{\rm f}_{\ell}>x^*_{\ell}$ 
	and $x_{\ell}^*\neq B_{\ell}/v_{\ell}$.
	Assume that ${\ell}\in \mathcal T_k\setminus \mathcal T_{k-1}$ for some $k\in\{1,2,\ldots,\tau\}$, 
	where $\mathcal T_0, \mathcal T_1, \ldots, \mathcal T_\tau$ is the chain of tight sets in Proposition~\ref{dropping}.
	By Proposition \ref{optimal}, there exists a set $H\subseteq H_{\ell}$ such that
	$x_{\ell}^*=f(H\cup \ell)-x^*(H)< B_{\ell}/v_{\ell}$.
	By the definition of $H_{\ell}$, 
	it must hold  
	$i\notin \{i_{1},i_{2},\ldots, i_{\tau}\}$ for each $i\in H_{\ell}\setminus \mathcal T_{k-1}$, which implies  
	$p^{\rm f}_i=B_i$ by Proposition~\ref{dropping}~(iii).
	Therefore, by $x^{\rm f}_{i}\geq p^{\rm f}_i/v_i$ (IR of buyers), we have
	\[
	x^{\rm f}(H\setminus \mathcal T_{k-1})\geq\sum_{i\in H\setminus \mathcal T_{k-1}}p^{\rm f}_i/v_i
	=\sum_{i\in H\setminus \mathcal T_{k-1}}B_i/v_i\geq x^*(H\setminus \mathcal T_{k-1}),
	\]
	where the last inequality holds by Proposition \ref{optimal}.
	Combining this with $x^{\rm f}_{\ell}>x^*_{\ell}$, we have 
	\begin{align*}
	x^*((H\cup \ell)\setminus \mathcal T_{k-1})<x^{\rm f}((H\cup \ell)\setminus \mathcal T_{k-1})\leq f((H\cup \ell)\cup \mathcal T_{k-1})-f(\mathcal T_{k-1})\leq f(H\cup \ell)-f(H\cap \mathcal T_{k-1}),
	\end{align*}
	where the second inequality holds by 
	feasibility constraint and the tightness of $\mathcal T_{k-1}$, and 
	the last inequality holds by $\ell\notin \mathcal T_{k-1}$ and the submodularity of $f$. 
	By $x^*(H\cup \ell)=f(H\cup \ell)$, we have 
	\[
	f(H\cup \ell)=x^*((H\cup \ell)\setminus \mathcal T_{k-1})+x^*(H\cap \mathcal T_{k-1})< f(H\cup \ell)-f(H\cap \mathcal T_{k-1})+x^*(H\cap \mathcal T_{k-1}).
	\]
	Therefore, we have $x^*(H\cap \mathcal T_{k-1})>f(H\cap \mathcal T_{k-1})$, 
	contradicting with the feasibility constraint.

	By Proposition \ref{optimal}, it holds $x^*_i\leq f(i)<B_i/v_i=\infty$ for any $i\in V$.
	By the contraposition of the first statement, we have
	$x^{\rm f}_i\leq x^*_i<B_i/v_i$.
	\end{proof}

\subsection{LW Guarantees}
	Now we provide the LW guarantee of the polyhedral clinching auction. 
	Let ${\rm LW}^{{\rm PCA}((N, M, E), v, \rho, B, s)}$ denote the LW of the allocation by ${\rm PCA}((N, M, E), v, \rho, B, s)$ and ${\rm LW}^{{\rm OPT}((N, M, E), v, \rho, B, s)}$ denote the optimal LW under the same input.
	We often denote them as ${\rm LW}^{\rm PCA}$ and ${\rm LW}^{{\rm OPT}}$, respectively, when the context is clear. 
	Also, let $v_{max}$ (resp. $v_{min}$) denote the maximum (resp. minimum) value of valuations for buyers, including virtual buyers. The main result of this section is the following:
	
	\begin{theorem}
	\label{LWofHS}
	If $\varepsilon\leq v_{min}^2/(v_{max}-v_{min})$, it holds ${\rm LW}^{\rm PCA}\geq  1/2 \ {\rm LW}^{\rm OPT}$.
	\end{theorem}
	
	Note that $\varepsilon$ is an increment of the price at each iteration in Algorithm \ref{HS_PCA}.
	The following example shows that the condition on $\varepsilon$ and the LW guarantee in Theorem \ref{LWofHS} are tight.
	\begin{example}
	\label{tightness_LWofHS}
	Consider a multi-unit auction consisting of two buyers and one seller without constraints on the allocation of goods.
	Buyer~1 has valuation $v_{min}+\varepsilon$ and an infinite budget.
	Buyer~2 has valuation $v_{max}$ and budget $v_{min}$.
	The seller offers one unit of a good and valuation $v_{min}$.
	Then, in the LW optimal allocation, according to Proposition \ref{optimal}, 
	buyer~1 obtains $1-v_{min}/v_{max}$ unit, and 
	buyer 2 obtains $v_{min}/v_{max}$ unit.
	Thus, ${\rm LW}^{\rm OPT}=v_{min}+(v_{min}+\varepsilon)\left(1-v_{min}/v_{max}\right)$.
	In Algorithm \ref{HS_PCA}, when the price clock of virtual buyer is equal to $v_{min}$, 
	she drops out of the auction.
	Similarly, when the price clock of buyer 1 is equal to $v_{min}+\varepsilon$, 
	she also drops out, and buyer 2 wins one unit of the good by paying $v_{min}$.
	Thus, we have ${\rm LW}^{\rm PCA}=\min(v_{max}, v_{min})=v_{min}$.
	If $\varepsilon>v_{min}^2/\left(v_{max}-v_{min}\right)$, then it holds 
	\begin{equation*}
	{\rm LW}^{\rm PCA}=v_{min} < \frac{1}{2}\left(v_{min}+(v_{min}+\varepsilon)
	\left(1-v_{min}/v_{max}\right)\right)=\frac{1}{2}\ {\rm LW}^{\rm OPT}.
	\end{equation*}
	Furthermore, if $\varepsilon=v_{min}^2/(v_{max}-v_{min})$, then it holds 
	\begin{equation*}
	{\rm LW}^{\rm PCA}=v_{min} = \frac{1}{2}\left(v_{min}+\left(v_{min}+\varepsilon\right)
	\left(1-v_{min}/v_{max}\right)\right)=\frac{1}{2}\ {\rm LW}^{\rm OPT}.
	\end{equation*}
	\end{example}

	Using the following theorem, we first prove Theorem \ref{LWofHS}. 
	This theorem provides a lower bound on the total payment for the mechanism, which is the first (non-trivial) guarantee of payments.	 
	\begin{theorem}
	\label{payment}
	If $\varepsilon\leq v_{min}^2/(v_{max}-v_{min})$, it holds 
	$\displaystyle\sum_{i\in N; x^{\rm f}_i> x^*_i}p^{\rm f}_i\geq
	\sum_{i\in N;x^{\rm f}_i \leq x^*_i}v_{i} (x_{i}^*-x^{\rm f}_{i}).$
	\end{theorem} 
\begin{proof}[Proof of Theorem \ref{LWofHS}]
	By Proposition \ref{x>xopt}, for each virtual buyer $i\in V$, it holds $x^{\rm f}_i\leq x^*_i$.
	Then, the set of buyers $\{i\in N; x^{\rm f}_i> x^*_i\}$ contains only non-virtual buyers.
	For each non-virtual buyer $i\in N\setminus V$, it follows from IR for buyers (Theorem \ref{properties_sec3} (i)) that $v_i x^{\rm f}_i \geq p^{\rm f}_i$. 
	By $B_i\geq p^{\rm f}_i$ (budget feasibility), it holds that $\displaystyle\min(v_i x^{\rm f}_i, B_i)\geq p^{\rm f}_i$. 
	Therefore, using Theorem \ref{payment}, we have 
	\begin{align}
	\label{1LW}
	{\rm LW}^{\rm PCA}\geq\sum_{i\in N; x^{\rm f}_i> x^*_i}\min(v_i x^{\rm f}_i, B_i)\geq 
	\sum_{i\in N; x^{\rm f}_i> x^*_i}p^{\rm f}_i\geq \sum_{i\in N;x^{\rm f}_i \leq x^*_i}v_{i} (x_{i}^*-x^{\rm f}_{i}).
	\end{align}
	
	For buyer $i\in N$ with $x^{\rm f}_i\leq x^*_i$, from Proposition \ref{optimal}, 
	it holds that $x^{\rm f}_i \leq x^*_i \leq B_i/v_i$, and thus 
	$\displaystyle\min(v_i x^{*}_i, B_i)=v_i x^{*}_i$ and $\displaystyle\min(v_i x^{\rm f}_i, B_i)=v_i x^{\rm f}_i$. 
	Then, we also have 
	\begin{align*}
	\label{2LW}
	\sum_{i\in N;x^{\rm f}_i \leq x^*_i}v_{i} (x_{i}^*-x^{\rm f}_{i})=
	\sum_{i\in N;x^{\rm f}_i \leq x^*_i}\left(\min(v_i x^*_i, B_i)-\min(v_i x^{\rm f}_i, B_i)\right) 
	\geq {\rm LW}^{{\rm OPT}}-{\rm LW}^{{\rm PCA}}.
	\end{align*}
	Combining this with (\ref{1LW}), we have 
	${\rm LW}^{\rm PCA}\geq \frac{1}{2}{\rm LW}^{\rm OPT}$.
\end{proof}

Theorem \ref{LWofHS} drastically extends the result by \citet{DL2014}, as compared in Table~\ref{compatison_DL2014}, and is also the first LW guarantee under polymatroid constraints. 
In their setting, Proposition~\ref{dropping} is accompanied by an additional property where $\mathcal T_k\setminus \mathcal T_{k-1} = i_k$ for $k\geq 2$ (in the case of $\tau\geq 2$), resulting in a simpler structure. This allows them to prove Theorem \ref{LWofHS} using only Proposition \ref{dropping}.
However, this property does not hold in our general setting, which requires us to consider Propositions \ref{optimal} and \ref{x>xopt}, and Theorem \ref{payment}. 
	\begin{table}[htbp]
	\centering
	\captionsetup{skip=5pt}
	\renewcommand{\arraystretch}{1.1}
	\caption{Comparison of LW guarantees in clinching auctions}
	\label{compatison_DL2014}
	\begin{tabular}{|c|c|c|}
	  \hline
	   & \textbf{\citet{DL2014}} & \textbf{Our Result} \\
	  \hline
	  Approximation Ratio & \multicolumn{2}{|c|}{1/2-approximation} \\
	  \hline \hline
	  \multirow{2}{*}{Mechanism} & Adaptive clinching auction & Polyhedral clinching auction \\
	  & \citep{BCMX2010} & \citep{HS2022} \\
	  \hline
	  The Number of Sellers& One & Multiple \\
	  \hline
	  Environment & Multi-units & Polymatroidal \\
	  \hline
	\end{tabular}
	\end{table}
Specifically, Theorem~\ref{payment} is highly non-trivial, and its proof is detailed in Section 8, with the key steps summarized below.
\begin{itemize}
\item In Section 8.1, we provide the simpler formula for $f_{x,d}$ expressed in (\ref{remnant}), 
where $x$ and $d$ are the number of allocated goods and demands of buyers, respectively, in an iteration of Algorithm~1.
\item In Section 8.2, we show that for the set of buyers $S$, the number of remaining goods that they can transact in the future is at least $\sum_{i\in S}(x^{*}_i-x_i)$, which will be sold at a higher price.
\item In Section 8.3, we construct an invariant inequality for future payments in the mechanism using {\it backward} induction, which means that the induction starts at the end of the auction.
\item In Section 8.4, we obtain a lower bound on the total payments by considering the initial step.
\end{itemize}

These analyses, which are the main technical contribution of this paper, might also be useful for investigating other properties of the mechanism, in particular with respect to payments.

\subsection{SW Guarantees}
	In Theorem \ref{LWofHS}, we provided the tight LW guarantee for the polyhedral clinching auction.
	By the definition of LW, the loss of LW is caused by either of 
	(i) allocating more goods than the budget divided by the valuation to buyers with high valuations, or 
	(ii) allocating goods to buyers with low valuations who have few goods in the LW optimal allocation.	
	The first statement of Proposition~\ref{x>xopt} implies that the loss of LW is caused by~(i).
The tendency to over-allocate the goods to buyers with high valuations means that the total sum of the valuations for the allocated goods becomes high. This motivates us to investigate the SW of the mechanism.
	
	Let the SW of ${\rm PCA}((N,M,E),v, \rho, B, s)$ be denoted by ${\rm SW}^{{\rm PCA}((N,M,E), v, \rho, B, s)}$, or simply by ${\rm SW}^{\rm PCA}$ when the context is clear. 
	We also provide the tight SW guarantee for the mechanism.

	\begin{theorem}
	\label{SW}
	It holds $\displaystyle {\rm SW}^{\rm PCA}=
	\sum_{i\in N}v_i x^{\rm f}_i \geq \sum_{i\in N}v_i x^{*}_i={\rm LW}^{\rm OPT}$.
	\end{theorem}
	If all buyers have sufficiently large budgets, 
	${\rm LW}^{{\rm OPT}}$ is exactly the optimal SW that is at least ${\rm SW}^{{\rm PCA}}$.
	Therefore, this SW guarantee is also tight. 
	This implies the high efficiency of the mechanism, since 
	the optimal LW can be seen as a reasonable benchmark of SW.

	\begin{proof}
	We show by mathematical induction that for each $k\in \{0,1,2,\ldots,\tau\}$, it holds 
	\begin{equation}
	\label{SW_induction2}
	\sum_{i\in \mathcal T_{k}} v_i x^{\rm f}_i \geq \sum_{i\in \mathcal T_{k}} v_i 
	x^{*}_i+v_{i_k} (x^{\rm f}\left(\mathcal T_{k})- x^{*}(\mathcal T_{k})\right).
	\end{equation}
	In the case of $k=0$,  we have (\ref{SW_induction2}) 
	since both sides are equal to $0$ by $\mathcal T_0=\emptyset$ (where we set $v_{i_0}:=v_{i_1}$).
	Suppose that the inequality holds for $k-1$. 
	By Proposition \ref{dropping} (i), it holds 
	$x^{\rm f}(\mathcal T_{k-1})=f(\mathcal T_{k-1})\geq  x^{*}(\mathcal T_{k-1})$ and $x^{\rm f}(\mathcal T_{k})=f(\mathcal T_{k})\geq x^{*}(\mathcal T_{k})$. Therefore, we have
	\begin{align*}
	\sum_{i\in \mathcal T_{k}} v_i (x^{\rm f}_i-x^*_i) &\geq v_{i_{k-1}} ( x^{\rm f}(\mathcal T_{k-1})-x^{*}(\mathcal T_{k-1}))+\sum_{i\in \mathcal T_{k}\setminus \mathcal T_{k-1}} v_{i} (x^{\rm f}_i- x^{*}_{i}) \\
	&\geq v_{i_{k}} ( x^{\rm f}(\mathcal T_{k-1})- x^{*}(\mathcal T_{k-1}))+\sum_{i\in \mathcal T_{k}\setminus \mathcal T_{k-1}} v_{i_{k}} (x^{\rm f}_i- x^{*}_{i})
	= v_{i_{k}} (x^{\rm f}(\mathcal T_\tau)-x^{*}(\mathcal T_\tau),
	\end{align*}
	where the first inequality holds by induction and 
	the second inequality holds by $v_{i}\geq v_{i_{k}}$ 
	for each $i\in \mathcal T_{k}$ due to Proposition \ref{dropping} (iii) 
	and $x^{\rm f}_i\geq x^{*}_{i}\ (i\in \mathcal T_{k}\setminus (\mathcal T_{k-1}\cup i_{k}))$ 
	due to the definition of $x^{*}$.
	Therefore, we have (\ref{SW_induction2}) for each $k\in \{0,1,2,\ldots,\tau\}$.
	Finally, since $\mathcal T_\tau=N$, it holds $x^{\rm f}(\mathcal T_\tau)=f(N)=x^{*}(\mathcal T_\tau)$ 
	by Theorem \ref{properties_sec3} (ii) and Proposition \ref{optimal}. 	
	Therefore, we have $\sum_{i\in N}v_i x^{\rm f}_i \geq \sum_{i\in N}v_i x^{*}_i$.
	\end{proof}

\section{Single-Sample Extension of Polyhedral Clinching Auctions}
	In this section, we extend the polyhedral clinching auction to an efficient single-sample mechanism 
	using the efficiency guarantees in Section 4. Notably, we are the first to address budget-constrained auctions in two-sided markets under the single-sample assumption.

\subsection{Single-Sample Assumptions}
	We consider the following assumption of sellers' valuations:
	\begin{assmp}
	\label{Sample}
	The valuation of seller $j\in M$ is drawn independently from a probability distribution $G_j$, 
	and the auctioneer has access to a single sample $\tilde{\rho}_j \sim G_j$ 
	from the distribution $G_j$ of each seller $j\in M$, 
	even though the auctioneer cannot access to the distributions $G:=\{G_j\}_{j\in M}$.
	\end{assmp}
	\begin{remark}
	This setting is more general than the Bayesian setting 
	since the auctioneer has no information on the distributions $G$, 
	and these single samples can be obtained if the auctioneer knows $G$. 
	This means that the impossibility theorem \citep{MS1983} still holds. 
	This setting lies between the Bayesian settings and the prior-free settings in terms of generality.
	\end{remark}
	
	Let $\tilde{\rho}:=\{\tilde{\rho}_j\}_{j\in M}$ denote the single samples of sellers' valuations.
	Then, the information available to the auctioneer is modified to 
	$\mathcal I\cup \tilde{\rho}$ and the input to our model is changed to $(\mathcal I, \tilde{\rho}, v, \rho, G)$.
	A single-sample mechanism $\tilde{\mathcal M}$ is a map from this information $\mathcal I\cup \tilde{\rho}$ to allocation $\mathcal A$.
	
	The definitions of desirable properties and efficiency objectives remain unchanged. 
	In contrast, efficiency is evaluated based on the expected value of the realization of valuations and samples.
	In the following, we fix the information on buyers and distributions $G$, and change the valuations and sample 
	values for sellers.
	If a single-sample mechanism $\tilde{\mathcal M}$ satisfies DSIC, then the expected values of LW and SW by the mechanism are denoted by 
	\[
	\mathbb E_{\rho, \tilde{\rho}\sim G}\ [{\rm LW}^{\tilde{\mathcal M}((N, M, E), \rho, \tilde{\rho}, s)}]\ \  {\rm and}\  \ 
	\mathbb E_{\rho, \tilde{\rho}\sim G}\ [{\rm SW}^{\tilde{\mathcal M}((N, M, E), \rho, \tilde{\rho}, s)}],
	\]
	respectively.
	We are interested in evaluating these expected values using the expected value of the optimal LW as a benchmark, 
	which is denoted by $\mathbb E_{\rho \sim G}$\ $[{\rm LW}^{{\rm OPT}((N, M, E), \rho, s)}]$.
	
	To avoid analyzing expectations, we devise the following trick that 
allows us to evaluate the efficiency using only pairwise inputs.
Let $\rho^a$ and $\rho^b$ be two sets of realized values from the distributions of sellers.
We consider two inputs to our model. 
In the first (resp. second) input, $\rho^a$ (resp. $\rho^b$) is used as samples, and $\rho^b$ (resp. $\rho^a$) is used as valuations. 
The remaining parts are common in these two inputs.
We evaluate the sum of the efficiency objectives achieved by a mechanism under these inputs 
by the sum of the optimal LW values under the same inputs.
The following lemma shows that if an approximation ratio is achieved 
under these pairwise inputs, the ratio is achieved in expectation.
It also holds when we replace the LW of the mechanism with its SW.	
	
	\begin{lemma}
	\label{Approx}
	Let $\tilde{\mathcal M}: \mathcal I\cup \tilde{\rho}\to \mathcal A$ be a single-sample mechanism that satisfies DSIC. 
	Let $\alpha$ be some positive constant with $\alpha\geq 1$.
	Suppose that it holds 
	\begin{equation}
	\label{assmp_sample}
	{\rm LW}^{\tilde{\mathcal M}((N, M, E), \rho^a, \rho^b, s)}+{\rm LW}^{\tilde{\mathcal M}((N, M, E), \rho^b, \rho^a, s)}\geq 
	\frac{1}{\alpha}({\rm LW}^{{\rm OPT}((N, M, E), \rho^a, s)}+{\rm LW}^{{\rm OPT}((N, M, E), \rho^b, s)})
	\end{equation}
	for arbitrary random variables $\rho^a, \rho^b$ with $\rho^a_j, \rho^b_j\sim G_j$ for each $j\in M$.
	Then, it holds 
	\[
	\mathbb E_{\rho, \tilde{\rho}\sim G}[{\rm LW}^{\tilde{\mathcal M}((N, M, E), \rho, \tilde{\rho}, s)}]\geq 
	\frac{1}{\alpha}\mathbb E_{\rho\sim G}[{\rm LW}^{{\rm OPT}((N, M, E), \rho, s)}].
	\]
	\end{lemma}
\begin{proof}
	The probabilities of $(\rho, \tilde{\rho})=(\rho^a, \rho^b)$ and 
	$(\rho, \tilde{\rho})=(\rho^b, \rho^a)$ are the same by symmetry, i.e., $\rho^a$ and $\rho^b$ are drawn independently from the same distribution $G$. 
	Then, we have
	\begin{eqnarray*}
	\mathbb E_{\rho^a, \rho^b\sim G}[{\rm LW}^{\tilde{\mathcal M}((N, M, E), \rho^a, \rho^b, s)}]&=&  
	\frac{1}{2}\mathbb E_{\rho^a, \rho^b\sim G}[{\rm LW}^{\tilde{\mathcal M}((N, M, E), \rho^a, \rho^b, s)}
	+{\rm LW}^{\tilde{\mathcal M}((N, M, E), \rho^a, \rho^b, s)}]\\
	&\geq&\frac{1}{2\alpha}\mathbb E_{\rho^a, \rho^b\sim G}[{\rm LW}^{{\rm OPT}((N, M, E), \rho^a, s)}
	+{\rm LW}^{{\rm OPT}((N, M, E), \rho^b, s)}]\\
	&=&\frac{1}{\alpha}\mathbb E_{\rho^a\sim G}[{\rm LW}^{{\rm OPT}((N, M, E), \rho^a, s)}],
	\end{eqnarray*}
	where the inequality holds by (\ref{assmp_sample}).
	\end{proof}

\subsection{Our Mechanism}
	Our single-sample mechanism $\mathbf M_{\rm sample}$ is obtained by the combination of 
	Algorithm \ref{HS_PCA} and the reduction method of~\citet{DFLLR2021}. 
	In this mechanism, sellers are first extracted and then 
	the polyhedral clinching auction with the extracted sellers are executed.
	For a set of sellers $M'\subseteq M$, let ${\rm PCA}|_{M'}(\rho')$ denote 
	${\rm PCA}((N, M', E'), v, \rho'_{M'}, B, s_{M'})$, where 
	$E':=\{ij\in E\colon j\in M'\}, 
	\ \rho'_{M'}:=\{\rho'_j\}_{j\in M'},\ s_{M'}:=\{s_j\}_{j\in M'}$.
	This corresponds to the execution of the polyhedral clinching auction 
	on the market consisting only of sellers in $M'$ when the bids of sellers are assumed to be $\rho'$.
	
	The full description of the mechanism $\mathbf M_{\rm sample}$ is given in Algorithm~\ref{First}. 
	\begin{algorithm}[htb]
	\caption{Our single-sample mechanism $\mathbf M_{\rm sample}((N, M, E), v, \rho', \tilde{\rho}, B, s)$}
	\label{First}           
	\begin{algorithmic}[1]
	  \STATE Extract the set of sellers $M':=\{j\in M: \tilde{\rho}_j \geq \rho'_j\}$.
	  \STATE Execute $PCA|_{M'}(\tilde{\rho})$ and obtain the allocation $(w, p, r)$ for the extracted market.
	  \STATE $w^{\rm f}_{ij}:=
	  \begin{cases}
	 w_{ij}\ \ {\rm if}\ j\in M'\\ 
	  0 \ \ \ \  \  {\rm o.w.}
	  \end{cases}(ij\in E)$, \ \ 
	  $p^{\rm f}_{i}:=p_{i}\ (i\in N)$,\ \ \ 
	  $r^{\rm f}_{j}:=
	  \begin{cases}
	 \tilde{\rho}_j w^{\rm f}(E_j)\ \ {\rm if}\ j\in M'\\ 
	  0 \quad\quad\quad \ \ \ \, {\rm o.w.}
	  \end{cases}(j\in M).$
	\end{algorithmic}
	\end{algorithm}
	Now we outline the mechanism. In line 1, sellers whose sample values are more than their bids are extracted. 
	In line~2, the polyhedral clinching auction is executed, where the market and constraints are restricted 
	to the extracted sellers and the bids of sellers are replaced with their sample values.
	Then, the allocation $(w,p,r)$ for the restricted market is obtained. 
	In line~3, this allocation is extended to the original market. 
	The allocation for the sellers not extracted in line 1 are all set to zero.  
	Also, the revenues of the extracted sellers are reduced so that the price per unit is equal to her sample value,
	which is needed for DSIC of sellers.\footnote{The reduced revenue from the sellers is assumed to be the auctioneer's revenue. We can randomly assign it to a seller. In this case, the mechanism satisfies SBB.} 
	Then, the final allocation $(w^{\rm f}, p^{\rm f}, r^{\rm f})$ is obtained.
	
In the rest of this section, we investigate the properties of this mechanism. We begin by focusing on the fundamental properties of the mechanism $\mathbf{M}_{\rm sample}$.

	\begin{theorem}
	\label{properties-1st}
	Mechanism $\mathbf M_{\rm sample}$ satisfies all of DSIC, IR, and WBB.
	\end{theorem}
	\begin{proof}
	In mechanism $\mathbf M_{\rm sample}$, the number of allocated goods and the payment of each buyer are equal to those obtained in line 2. 
	Then, by Theorem \ref{properties_sec3} (i), 
	mechanism $\mathbf M_{\rm sample}$ satisfies DSIC and IR for buyers.
	Moreover, each seller's bid is only used to determine whether the seller 
	participates in the auction, and the revenue is independent of the bid. 
	If the valuation $\rho_j$ of seller $j$ is greater than $\tilde{\rho}_j$, the seller prefers not to participate 
	because she is unwilling to sell her goods at the price $\tilde{\rho}_j$.
	If $\rho_j$ is equal to $\tilde{\rho}_j$ or less, she is willing to sell her goods at $\tilde{\rho}_j$. 
	In both cases, the best choice for seller $j$ is achieved by truthful bidding $\rho'_j=\rho_j$,  
	and she does not sell her goods at a price below her valuation.
	Thus, the mechanism also satisfies DSIC and IR for sellers.
	
	Let $(w,p,r)$ be the allocation obtained in line 2.
	Then, the total payment $\sum_{i\in N} p_i$ is equal to the total revenue 
	$\sum_{j\in M'} r_j$ by SBB of the polyhedral clinching auction (Theorem~\ref{properties_sec3} (i)). 
	By $r_j\geq \tilde{\rho}_j w(E'\cap E_j)\ (j\in M')$ from IR for sellers (Theorem~\ref{properties_sec3} (i)), 
	we have 
	\begin{equation*}
	\sum_{i\in N} p^{\rm f}_i=\sum_{i\in N} p_i=\sum_{j\in M'} r_j \geq \sum_{j\in M}\tilde{\rho}_j 
	w(E'\cap E_j)=\sum_{j\in M}\tilde{\rho}_j w^{\rm f}(E_j)=\sum_{j\in M}r^{\rm f}_j. 
	\end{equation*}
	Thus, the mechanism satisfies WBB.
	\end{proof}

We then show that the reduction method of \citet{DFLLR2021}, which uses the VCG mechanism and provides a strong SW guarantee, can also provide a strong LW guarantee when applied to the polyhedral clinching auction under budget constraints.

	To analyze the efficiency guarantees of the mechanism $\mathbf M_{\rm sample}$, we use pairwise inputs. Let $\rho^a$ and $\rho^b$ be arbitrary random variables with $\rho^a_j, \rho^b_j \sim G_j$ for each $j\in M$.
	We denote ${\rm LW}^{\tilde{\mathcal M}((N, M, E), \rho^a, \rho^b, s)}$ 
	and ${\rm LW}^{{\rm OPT}((N, M, E), \rho^a, s)}$
	by ${\rm LW}^{\tilde{\mathcal M}( \rho^a, \rho^b)}$ and ${\rm LW}^{{\rm OPT}(\rho^a)}$,
	respectively, when it is clear from the context.
	Our goal is the following efficiency guarantees:
	\begin{theorem}
	\label{efficiency-1st}
	In mechanism $\mathbf M_{\rm sample}$, the following holds:
	\begin{itemize}
	\item[(i)] It holds ${\rm LW}^{\mathbf M_{\rm sample}(\rho^a, \rho^b)}+{\rm LW}^{\mathbf M_{\rm sample}(\rho^b, \rho^a)} \geq \frac{1}{4}({\rm LW}^{{\rm OPT}(\rho^a)}+ {\rm LW}^{{\rm OPT}(\rho^b)})$.
	\item[(ii)] It holds ${\rm SW}^{\mathbf M_{\rm sample}(\rho^a, \rho^b)}+{\rm SW}^{\mathbf M_{\rm sample}(\rho^b, \rho^a)} \geq \frac{1}{2}({\rm LW}^{{\rm OPT}( \rho^a)}+ {\rm LW}^{{\rm OPT}(\rho^b)})$.
	\end{itemize}
	\end{theorem}
	Note that these efficiency guarantees are half of Theorems \ref{LWofHS} and~\ref{SW}, 
	which is consistent with Theorem 2 of \citet{DFLLR2021}. 
	The efficiency loss results from the deletion of sellers in line 1 of Algorithm~\ref{First}.	
	By Lemma \ref{Approx}, we can see that the mechanism achieves LW of more than 1/4 of the optimal LW, and achieves SW of more than 1/2 of the optimal LW, in expectation. Then, by Theorem~\ref{properties-1st}, we have shown all the properties in Main Theorem. 
	Moreover, the following example shows that the above efficiency guarantees are tight.
	\begin{example}
	\label{tight_ex}
	Consider a multi-unit auction with two buyers and one seller.
	Buyer~1 has valuation $1$ and infinite budget, and 
	buyer~2 has valuation $k\ (k\geq 1)$ and budget 1.
	The seller owns one unit of a good.
	For an arbitrary small positive number $\Delta\ll 1$, we consider two cases: 
	$(\rho_1, \tilde{\rho}_1)=(\Delta, 2\Delta)$ and $(\rho_j, \tilde{\rho}_j)=(2\Delta, \Delta)$.
	Then, in both cases, the optimal LW is $2-1/k$ because $\Delta\ll 1$.
	
	If $(\rho_1, \tilde{\rho}_1)=(\Delta, 2\Delta)$, then the seller participates in the auction and sells goods, and 
	when the price of buyer 1 reaches 1, all goods are allocated to buyer 2.
	In this case, the LW is $\min(k, 1)=1$, and the SW is $k$.
	If $(\rho_1, \tilde{\rho}_1)=(2\Delta, \Delta)$, then the seller does not participate in the auction.
	In this case, the LW and SW are $2\Delta$.
	Thus, it holds 
	\[
	{\rm LW}^{\mathbf M_{\rm sample}(2\Delta,\Delta)}+{\rm LW}^{\mathbf M_{\rm sample}(\Delta,2\Delta)}=1+2\Delta\leq \frac{1+2\Delta}{2(2-1/k)}({\rm LW}^{{\rm OPT}(2\Delta)}+{\rm LW}^{{\rm OPT}(\Delta)}).
	\]
	Taking $k\to\infty$ and $\Delta\to 0$, we have $\frac{1+2\Delta}{2(2-1/k)}\to \frac{1}{4}$.
	Additionally, it holds that 
	\[
	{\rm SW}^{\mathbf M_{\rm sample}(2\Delta,\Delta)}+{\rm SW}^{\mathbf M_{\rm sample}(\Delta,2\Delta)}=k+2\Delta\leq \frac{k+2\Delta}{2(2-1/k)}({\rm LW}^{{\rm OPT}(2\Delta)}+{\rm LW}^{{\rm OPT}(\Delta)}).
	\]
	Taking $k\to 1$ and $\Delta\to 0$, we have $\frac{k+2\Delta}{2(2-1/k)}\to \frac{1}{2}$.
	Therefore, our efficiency guarantees are tight.
	\end{example}

	Now we prove Theorem \ref{efficiency-1st}.
	Let ${\rm LW}^{{\rm PCA}|_{M'}(\rho')}$ denote the LW value obtained by the execution of ${\rm PCA}|_{M'}(\rho')$.
	The following lemma provides the relationship 
	between the LW of the mechanism $\mathbf M_{\rm sample}$ and the LW of Algorithm \ref{HS_PCA}. 
	Define $M_a:=\{j\in M: \rho^a_j \geq \rho^b_j\}$ and $M_b:=\{j\in M: \rho^a_j \leq\rho^b_j\}$.
	Remark that the same argument holds if we replace LW with SW.
	
	\begin{lemma}
	\label{relation_PCA}
	In mechanism $\mathbf M_{\rm sample}$, it holds 
	\begin{align*}
	{\rm LW}^{\mathbf M_{\rm sample}(\rho^a, \rho^b)}+{\rm LW}^{\mathbf M_{\rm sample}(\rho^b, \rho^a)}
	&\geq{\rm LW}^{{\rm PCA}|_{M_a}(\rho_a)}+{\rm LW}^{{\rm PCA}|_{M_b}(\rho_b)}.
	\end{align*}
	\end{lemma}
	\begin{proof}
	Let $x^a:=\{x^a_j\}_{j\in M_a}$ (resp. $x^b:=\{x^b_j\}_{j\in M_b}$) denote the number of unsold goods of sellers in 
	${\rm PCA}|_{M_a}(\rho_a)$ 
	(resp. ${\rm PCA}|_{M_b}(\rho_b))$ in line 2 of Algorithm~\ref{First}, which
	also represents the number of allocated goods to virtual buyers corresponding to the sellers in $M_a$ (resp. $M_b$).
	
	In $\mathbf M_{\rm sample}((N,M,E), v, \rho^a, \rho^b, B, s)$, 
	the allocation for buyers (including virtual buyers) is the same as that of ${\rm PCA}|_{M_b}(\rho_b)$.
	However, the valuations of sellers in $M_b$ is actually $\rho^a$, 
	not $\rho^b$ used in the polyhedral clinching auction.
	Moreover, sellers in $M\setminus M_b$ do not participate in the auction, 
	and thus their goods are unsold at the end of the auction.
	Thus, the LW of the allocation by mechanism $\mathbf M_{\rm sample}((N,M,E), v, \rho^a, \rho^b, B, s)$ is expressed as 
	\begin{equation}
	\label{eq71}
	{\rm LW}^{\mathbf M_{\rm sample}(\rho^a, \rho^b)}= {\rm LW}^{{\rm PCA}|_{M_b}(\rho_b)}
	-\sum_{j\in M_b}(\rho^b_j-\rho^a_j)x^{b}_{j}+\sum_{j\in M\setminus M_b} \rho^a_j s_j.
	\end{equation}
	Similarly, by swapping $\rho^a$ and $\rho^b$, we also have  
	\begin{equation}
	\label{eq72}
	{\rm LW}^{\mathbf M_{\rm sample}(\rho^b, \rho^a)}= {\rm LW}^{{\rm PCA}|_{M_a}(\rho_a)} 
	-\sum_{j\in M_a}(\rho^a_j-\rho^b_j)x^{a}_{j}+\sum_{j\in M\setminus M_a} \rho^b_j s_j.
	\end{equation}
	By $M_a=(M\setminus M_b)\cup \{j\in M: \rho^a_j=\rho^b_j\}$ and 
	$M_b=(M\setminus M_a)\cup \{j\in M: \rho^a_j=\rho^b_j\}$, we~have 
	\begin{align*}
	\sum_{j\in M_b}(\rho^b_j-\rho^a_j)x^{b}_{j}=\sum_{j\in M\setminus M_a}(\rho^b_j-\rho^a_j)x^{b}_{j}\ \ \ {\rm and}\ \ \ 
	\sum_{j\in M_a}(\rho^a_j-\rho^b_j)x^{a}_{j}=\sum_{j\in M\setminus M_b}(\rho^a_j-\rho^b_j)x^{a}_{j}.
	\end{align*}
	Combining this with the supply constraints $x^{b}_{j}\leq s_j$ and $x^{a}_{j}\leq s_j$ for each $j\in M$, we have 
	\begin{align}
	\label{eq73}
	-\sum_{j\in M_b}(\rho^b_j-\rho^a_j)x^{b}_{j}+\sum_{j\in M\setminus M_b} \rho^a_j s_j-
	\sum_{j\in M_a}(\rho^a_j-\rho^b_j)x^{a}_{j}+\sum_{j\in M\setminus M_a} \rho^b_j s_j\geq 0.
	\end{align}
	Summing \eqref{eq71}, \eqref{eq72}, and \eqref{eq73}, we obtain
	\[
	\displaystyle
	{\rm LW}^{\mathbf M_{\rm sample}(\rho^a, \rho^b)}+{\rm LW}^{\mathbf M_{\rm sample}(\rho^b, \rho^a)}\geq {\rm LW}^{{\rm PCA}|_{M_a}(\rho_a)}+{\rm LW}^{{\rm PCA}|_{M_b}(\rho_b)}.
	\]
	\end{proof}

	Let ${\rm LW}^{{\rm OPT}|_{M'}(\rho)}$ denote 
	the optimal LW value in the market where sellers are restricted to $M'\subseteq M$ and have valuations $\rho|_{M'}$.
	We also use the following property on the optimal LW.
	\begin{lemma}
	\label{relation_optimal}
	For the optimal value of LW, it holds that 
	\begin{align*}
	{\rm LW}^{{\rm OPT}|_{M_a}(\rho^a)}
	+{\rm LW}^{{\rm OPT}|_{M_b}(\rho^b)}
	\geq
	\frac{1}{2}({\rm LW}^{{\rm OPT}(\rho^a)}+{\rm LW}^{{\rm OPT}(\rho^b)}).
	\end{align*}
	\end{lemma}
	The proof is provided in Appendix C.
	Using this lemma, we provide the proof of Theorem \ref{efficiency-1st}.
	\begin{proof}[Proof of Theorem \ref{efficiency-1st}]
	(i): From Theorem \ref{LWofHS}, we have the following:
	\begin{align}
	\label{PCA1}
	{\rm LW}^{{\rm PCA}|_{M_a}(\rho_a)}\geq \frac{1}{2} 
	{\rm LW}^{{\rm OPT}|_{M_a}(\rho^a)}\quad {\rm and}\quad
	{\rm LW}^{{\rm PCA}|_{M_b}(\rho_b)}\geq \frac{1}{2} 
	{\rm LW}^{{\rm OPT}|_{M_b}(\rho^b)}.
	\end{align}
	Combining this with Lemma \ref{relation_PCA}, we obtain the following:
	\begin{align*}
	{\rm LW}^{\mathbf M_{\rm sample}(\rho^a, \rho^b)}+{\rm LW}^{\mathbf M_{\rm sample}(\rho^b, \rho^a)}
	&\geq \frac{1}{2}({\rm LW}^{{\rm OPT}|_{M_a}(\rho^a)}+
	{\rm LW}^{{\rm OPT}|_{M_b}(\rho^b)})\\
	&\geq \frac{1}{4}({\rm LW}^{{\rm OPT}(\rho^a)}+{\rm LW}^{{\rm OPT}(\rho^b)}),
	\end{align*}
	where the second inequality holds from Lemma \ref{relation_optimal}.
	
	(ii): In the proof of property (i), we can replace (\ref{PCA1}) with the following:
	\begin{align*}
	{\rm SW}^{{\rm PCA}|_{M_a}(\rho_a)}\geq
	{\rm LW}^{{\rm OPT}|_{M_a}(\rho^a)}\quad {\rm and}\quad
	{\rm SW}^{{\rm PCA}|_{M_b}(\rho_b)}\geq
	{\rm LW}^{{\rm OPT}|_{M_b}(\rho^b)},
	\end{align*}
	which holds by Theorem \ref{SW}. 
	By Lemmas \ref{relation_PCA} (for SW) and \ref{relation_optimal}, property (ii) also holds.
	\end{proof}
	
\section{Further Extensions}
In this section, we extend our results to the case of polymatroidal constraints and indivisible goods.
\subsection{Polymatroid Constraint}
Our results can be easily extended to polymatroid contraints, 
which is known to the most generalized setting handled in clinching auctions.
In Appendix B, for readers unfamiliar with polymatroid theory, we briefly summarize the basics of the theory relevant to this study.

Instead of the supply constraint $w(E_j)\leq s_j\ (j\in M)$, 
we consider the scenario where the transaction 
$w|_{E_j}$ of each seller $j$ is constrained by a polymatroid $P_j$ on $E_j$, i.e., 
$w|_{E_j}\in P_j\ (j\in M)$.\footnote{In this setting, polymatroids are assumed to be public due to the impossibility theorem 
by \citet{FLSS2011}.} 
The monotone submodular function that defines the polymatroid $P_j$ is denoted by $f_j$. 
Then, the supply $s_j$ of seller $j$ is represented by $f_j(E_j)$. 
As discussed in \citet{HS2022}, with the following modifications, Algorithm~\ref{HS_PCA} outputs a feasible allocation under the polymatroid constraint. 
Notably, all of our technical results can be extended directly to this setting after replacing $f(S):=\sum_{j\in M_S}s_j$ with $f(S):=f_j(E_j\cap E_S)$. Then, all the proofs in Sections 3 to 5, as well as Section 8, hold without modification. This means that even under polymatroid constraints, our single-sample mechanism is also applicable and the efficiency guarantee of Theorem \ref{efficiency-1st} holds immediately.

\begin{figure}[htbp]
	\centering
	\includegraphics[width=90mm]{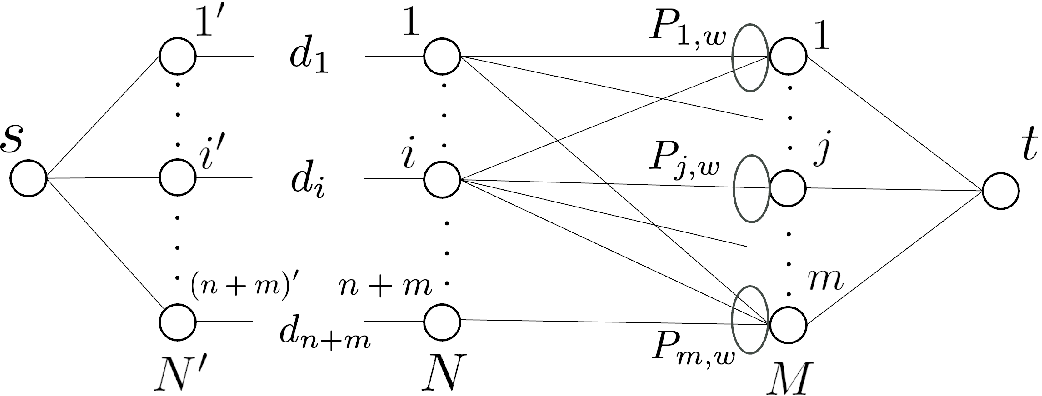}
	\caption{The polymatroidal network $\mathcal N$.}
	\label{polymatroidal}
	\end{figure}

\paragraph{The Inputs of Algorithms.} Instead of $s$, 
we use $P:=\{w\in \mathbf R^{E}_+ \colon w|_{E_j}\in P_j~(j\in~M)\}$. Also, in Algorithm \ref{First}, 
$P_{M'}:=\{w\in \mathbf R^{E'}_+ \colon w|_{E_j}\in P_j\ (j\in M')\}$ is used instead of~$s_{M'}$.
\paragraph{Preprocessing (line 1).} Along with the modifications in Section 3.1, we extend $f_j\, (j\in M)$ to 
	\[
	f_j (F):=
	\begin{cases}
	f_j (E_j)          \qquad  {\rm if}\ (n+j)j\in F, \\
	f_j (F)       \qquad\  {\rm otherwise},
	\end{cases}
	 (F \subseteq E_j \cup \{(n+j)j\}).
	\]
	The polymatroid $P_j$ is also extended according to $f_j$.
\paragraph{Constructing the Network $\mathcal N$ in the Clinching Procedure (line 4).} 
	For polymatroids $\{P_j\}_{j\in M}$ and the transaction vector $w$, and the demand vector $d$, 
	we construct the polymatroidal network $\mathcal N$ as follows:
	Suppose that each edge in $E$ is directed from $N$ to $M$. 
	For each $i\in N$, consider its copy $i'$ and connect 
	it with a directed edge $i'i$. 
	Add a source node $s$ and connect it with 
	each $i'\in N'$ through a directed edge $si'$. 
	Also, add a sink node $t$ and connect each $j\in M$ 
	to it with a directed edge $jt$. The edge capacities are defined by 
	\begin{align*}
	{\rm cap}(e):=
	\begin{cases}
	d_i&{\rm if}\ e=i'i\ {\rm for}\ i\in N,\\
	\infty&{\rm otherwise}.
	\end{cases}
	\end{align*}
	For the polymatroid $P_j$ and the transaction vector $w$, 
	define a polytope $P_{j,w}$ on $E_j$ 
	that represents the feasible future transactions of seller $j$ 
	in terms of polymatroid constraints~by 
	\[
	P_{j,w}:=\{w'\in \mathbf R^{E_j}_+ \colon w|_{E_j}+w'\in P_j\}.
	\]
	This polytope is obtained by contraction by a vector $w|_{E_j}$ and is therefore a polymatroid (see Appendix~B).
	For seller $j\in M$, we set the polymatroid 
	with ground set $\delta^-_j=E_j$ as $P_{j,w}$, 
	and the one with ground set 
	$\delta^+_j$ as defined by the edge capacities. 
	The polymatroids corresponding to other nodes are 
	also defined by the edge capacities.
	The resulting network is illustrated in Figure~\ref{polymatroidal}.
	
	Now we consider the maximum flow value in $\mathcal N$. Define a function 
	$g_{w,d}: \mathbf R^E_{+}\to \mathbf R_+$ by 
	\[
	\displaystyle g_{w,d}(F):=\max_{\psi:\, {\rm flow\, in\,}\mathcal N}\psi(F)\quad (F\subseteq E).
	\]
	By solving the maximum polymatroidal flow problem, 
	\citet{HS2022} showed that for each $F\subseteq E$, the value $g_{w,d}(F)$ can be computed in polynomial time,  
	provided that the value oracles of $f_j\ (j\in M)$ are given. In particular, $g_{w,d}(F)$ is expressed as:
	\begin{equation}
	\label{gwd}
	g_{w,d}(F)=\min_{S\subseteq N_F}\{\min_{F'\supseteq E_{S}\cap F}
	\{\sum_{j\in M}f_j(F'\cap E_j)-w(F')\}+d(N_F\setminus S)\}\quad (F\subseteq E),
	\end{equation}
	where $N_F$ denotes the set of nodes in $N$ incident to~$F$.
	They also showed that the clinching polytope $P^{i}_{w,d}$ is a polymatroid and the submodular function 
	$g^i_{w,d}:2^{E_i}\to\mathbf R_+$ that defines $P^{i}_{w,d}$ is given by 
	$g^i_{w,d}(F)=g_{w,d}(E_{N\setminus i}\cup F)-g_{w,d}(E_{N\setminus i})$. 
	This result was used to prove Theorem \ref{clinch_sec3}.

\subsection{Indivisible Goods}

\citet{HS2023} proposed the polyhedral clinching auction for indivisible goods in one-sided markets. They extended our efficiency guarantees by integrating our techniques with their original contributions to address the challenges posed by indivisibility. Their results show not only the effectiveness of our methods in Section 8, but also the potential to extend our results to the indivisible setting, where $w$ must be an integer vector, given $s_j\in \mathbf Z_+\ (j\in M)$.
Now we explain how our results can be extended to this setting by incorporating the multidimensional clinching of  \citet{HS2022} into the mechanism developed by \citet{HS2023}. As in Sections 3 and 4, we first consider the setting of  truthful sellers, i.e., $\rho_j=\rho'_j\ (j\in M)$.

	Now we propose the indivisible extension of Algorithm \ref{HS_PCA}, 
	where the full description is given in Algorithm~\ref{Indivisible3}. 
	The execution of the mechanism is summarized as follows: 
	In line 1, the preprocessing of Section~3.1 is again performed to construct 
	a modified market. Thereafter, the process is similar to the mechanism of \citet{HS2023} 
	although the allocation is changed from $(x,\pi)$ to $(w,p,r)$.
	As in Algorithm \ref{HS_PCA}, 
	the price clock $c\in \mathbf R_{+}$ represents a transaction price per unit and 
	gradually increases from zero.
	For the current price $c$, the demands $d:=(d_i)_{i\in N}$
	are determined.
	At the beginning of an iteration, the price clock $c$ is increased until 
	there appears a buyer $k$ with $c=v'_k$ or $d_k=(B_k-p_k)/c$. 
	Then, the demands of these buyers are updated and 
	Algorithm~\ref{HS_clinch} is directly applied in lines 8 and 12.
	After processing both cases of demand updates, 
	the next iteration is performed if there exists a buyer with a positive demand, and otherwise, 
	the mechanism terminates. 
	As in the divisible case, in the analysis, we use ($w^{\rm f}, p^{\rm f}, r^{\rm f}$) in line 15 as the final allocation.
	
	\begin{algorithm}[htb]
	\caption{Indivisible Extension of Algorithm \ref{HS_PCA}: ${\rm PCA_{ind}}((N,M,E),v', \rho', B, s)$} 
	\label{Indivisible3}
	\begin{algorithmic}[1]
	 \STATE Add virtual buyers $ V$ and construct a modified market $(N, M, E)$. 
	  \STATE $d_i:=\sum_{j\in M_i}s_j+1 \  (i\in N)\ {\rm and}\ c:=0$.\\
	  \STATE  $p_i:=0,\ (i\in N)$,\ $r_j:=0\ (j\in M)\ {\rm and} \ w_{ij}:=0\ (ij\in E)$.
	  \WHILE{active buyers exist}
	  \STATE Increase $c$ until there appears an active buyer 
	  $k$ with $v'_{k}=c$ or $d_{k}=\frac{B_{k}-p_{k}}{c}$.
	  \WHILE{$\exists$ active buyer $k$ with $v'_{k}=c$}
	  \STATE Pick such a buyer $k$ and let $d_{k}:=0$.
	  \STATE Clinching$(s, (w, p, r), d, c)$.
	  \ENDWHILE
	  \WHILE{$\exists$ active buyer $k$ with $d_k=\frac{B_k-p_k}{c}$}
	  \STATE Pick such a buyer $k$ and let $d_{k}:=d_{k}-1$.
	  \STATE Clinching$(s, (w, p, r), d, c)$.
	  \ENDWHILE
	 \ENDWHILE
\STATE  $w^{\rm f}_{ij}:=w_{ij}\ (ij\in E)$,\ $p^{\rm f}_i:=p_i\ (i\in N)$,\ {\rm and} \  $r^{\rm f}_j:=r_j\ (j\in M)$.
	\STATE  Output the final allocation after cancelling all transactions between virtual buyers~and~sellers.
	\end{algorithmic}
	\end{algorithm}
	
	\begin{rem}
	The total sum of demands is decreased by at least one in each iteration.
	Then, Algorithm~\ref{Indivisible3} terminates after at most $(n+m)\sum_{j\in M}s_j+(n+m)$ iterations.
	Since each iteration can be computed in polynomial time,  
	the entire procedure can also be computed in polynomial time.
	\end{rem}

We first verify that the final allocation satisfies the integer constraint. Given that $s_j\in \mathbf Z_+$ for each $j$, if both $w$ and $d$ are integer vectors, the functions $f_{x,d}$ and $g_{w,d}$ in (\ref{gwd}) are also integer-valued. 
This implies that, in Algorithm~\ref{HS_clinch}, the clinching polytope becomes an integer polymatroid, where a maximal integer vector $\xi_i\in P^{i}_{w,d}\ (i\in N)$ can be computed by the greedy procedure. Then, $\xi_i$ is also an integer vector and thus both $w$ and $d$ remain integer vectors.
Since $w$ and $d$ are integer vectors at the beginning, $f_{x,d}$ and $g_{w,d}$ remain integer-valued throughout. 
Thus, the following holds:
\begin{lemma} 
\label{integer-clinching2} 
Throughout Algorithm \ref{Indivisible3}, all vectors $w$, $d$, and $\xi_i\ (i\in N)$ remain integer. 
\end{lemma}

	The efficiency guarantees can be imported from the results of \citet{HS2023}. 
	To see this point, consider a market, called \textit{reduced one-sided market}, which is constructed by the preprocessing in Section 3.1 and aggregating all sellers into one. Then, the mechanism of \citet{HS2023} is applicable to this market and let $(\tilde{x}^{\rm f},\tilde{\pi}^{\rm f})$ be the final allocation, 
where $\tilde{x}^{\rm f}\in \mathbf Z^N_+$ and $\tilde{\pi}^{\rm f}\in \mathbf R^N_+$ are the total number of allocated goods and the payment of buyers, respectively.
	Then, the following relations hold between $(\tilde{x}^{\rm f},\tilde{\pi}^{\rm f})$ and 
	$(w^{\rm f}, p^{\rm f}, r^{\rm f})$ under the same input.
	\begin{theorem}
	\label{relation}
	For each $i\in N$, it holds $\tilde{x}^{\rm f}_i=w^{\rm f}(E_i)$ and $\tilde{\pi}^{\rm f}_i=p^{\rm f}_i$ 
	\end{theorem}
	This property is the same as Theorem 7 of \citet{HS2022}, where 
	the proof can directly be adapted in this indivisible setting.
	It implies that all properties regarding the final allocation of buyers 
	for this reduced one-sided market also hold in Algorithm~\ref{Indivisible3}.
	Since \citet{HS2023} extended our Theorems \ref{LWofHS} and \ref{SW} to their indivisible setting, 
	these theorems also hold for Algorithm~\ref{Indivisible3}.
	Remark that they showed that these guarantees are also tight in the indivisible case.

	By Theorem \ref{relation}, DSIC and IR for buyers also hold. 
	Since Theorem \ref{clinch_sec3} (iii) implies IR for sellers and 
	SBB holds trivially, this mechanism satisfies all of DSIC for buyers, IR, and SBB.
	Therefore, all the properties needed for the single-sample extension hold in this mechanism.

Using Algorithm~\ref{Indivisible3} instead of Algorithm~\ref{HS_PCA}, Algorithm~\ref{First} can easily be extended to the indivisible setting. Since all the proofs of Section 5 remain valid, Main Theorem also holds. While this extension is technically achieved using the method developed by \citet{HS2022},  
it is particularly beneficial in showing that the same theoretical results can be obtained in both divisible and indivisible cases, since indivisibility often raises some difficulties.
Combined with the approach in Section 6.1, we can also address both polymatroidal and indivisible settings. Now we present an example corresponding to the indivisible case of Example 5.6.

	\begin{example}
	Consider a multi-unit auction with two buyers, and one seller.
	The seller owns $k$ units of a good. 
	Buyer 1 has valuation $1$ and budget $k$, and 
	buyer 2 has valuation $k\ (k\geq 1)$ and budget $k$.
	For an arbitrary small positive number $\Delta\ll 1$, we consider two cases: 
	$(\rho_1, \tilde{\rho}_1)=(\Delta, 2\Delta)$ and $(\rho_j, \tilde{\rho}_j)=(2\Delta, \Delta)$.
	In both cases, the optimal LW is $2k-1$ because $\Delta\ll 1$.
	
	If $(\rho_1, \tilde{\rho}_1)=(\Delta, 2\Delta)$, then the seller participates in the auction and sells her good, and when the price is 1, all the goods are allocated to buyer 2.
	In this case, the LW is $\min(k, k)=k$, and the SW is $k^2$.
	If $(\rho_1, \tilde{\rho}_1)=(2\Delta, \Delta)$, then the seller does not participate in the auction.
	In this case, the LW and SW are $2\Delta$.
	Thus, it holds 
	\[
	{\rm LW}^{\mathbf M_{\rm sample}(2\Delta,\Delta)}+{\rm LW}^{\mathbf M_{\rm sample}(\Delta,2\Delta)}=
	k+2\Delta\leq \frac{k+2\Delta}{2(2k-1)}({\rm LW}^{{\rm OPT}(2\Delta)}+{\rm LW}^{{\rm OPT}(\Delta)}).
	\]
	Taking $k\to\infty$ and $\Delta\to 0$, 
	we have $\frac{1+2\Delta/k}{2(2-1/k)}\to \frac{1}{4}$.
	Additionally, it holds that 
	\[
	 {\rm SW}^{\mathbf M_{\rm sample}(2\Delta,\Delta)}+ {\rm SW}^{\mathbf M_{\rm sample}(\Delta,2\Delta)}=k^2+2\Delta\leq \frac{k^2+2\Delta}{2(2k-1)}({\rm LW}^{{\rm OPT}(2\Delta)}+{\rm LW}^{{\rm OPT}(\Delta)}).
	\]
	Taking $k\to 1$ and $\Delta\to 0$, we have $\frac{k^2+2\Delta}{2(2k-1)}\to \frac{1}{2}$.
	Therefore, our efficiency guarantees are also tight in the indivisible settings.
	\end{example}

\subsection{Application: Ad Slot Auctions}
The extensions allow us to address practical applications of ad slot allocation, proposed in \citet{HS2022}, even under the single-sample assumption.\footnote{These are also two-sided extensions of the ad auctions in \citet{BHLS2015} and~\citet{GMP2015}.} We can handle these situations not only when ad slots are indivisible but also when they can be divided by time or other criteria.

Consider an ad exchange market where ad slots for certain time periods are allocated to advertisers in advance. Suppose that each publisher $j$ offers ad slots on their website, consisting of pages $\{1, 2,\ldots, \beta_j\}$, where each page $k\in \{1, 2, \ldots, \beta_j\}$ contains $t^k_j\in \mathbf Z_+$ slots. We impose a restriction so that each advertiser purchases at most one slot per page, ensuring that the same ad does not appear multiple times on the same page, which may be preferable for advertisers. This restriction can be implemented by defining the submodular function $f_j$ for seller $j$ as:
\[
f_j(F_j)=\sum_{k=1}^{\beta_j}\min(t^k_j, |F_j|)\quad (F_j\subseteq E_j).
\]
In \citet{HS2022}, more sophisticated methods have also been proposed, where impressions are allocated first, followed by the assignment of ad slots based on these impressions. This approach takes into account differences in the valuation of individual ad slots.

\section{Discussions}
In the previous sections, we established the efficiency guarantees of the polyhedral clinching auction and extended it to an efficient single-sample mechanism. This study is the first to address budget-constrained auctions in two-sided markets under the single-sample assumption. Moreover, our results hold under polymatroid constraints and apply to both divisible and indivisible goods, enabling us to address practical applications in display advertising. As shown in Examples 5.6 and~6.3, all efficiency guarantees for our mechanism are tight. In the following, we discuss whether the best possible guarantees are achieved and outline some directions for future research.

We begin by examining a bilateral trade involving a buyer and a seller offering one unit of a good. \citet{DFLLR2021} showed that any single-sample mechanism, where the auctioneer has access only to the seller's sample value, cannot achieve better than a 1/2-approximation to the optimal SW while satisfying DSIC, IR, and BB. 
Even when the good is divisible, the result of \citet{LRW2023} implies that this bound still holds if all transaction prices are independent of bids. These results suggest that our SW guarantee in Theorem~\ref{efficiency-1st} achieves the best possible bound. Since the optimal LW equals the optimal SW when the buyer has an infinite budget, the upper bound on the approximation ratio for LW is also at most 1/2. In Appendix~A, we further show that in bilateral trade (including the divisible case), our mechanism achieves a 1/2-approximation to the optimal LW, which is also the best possible.

In more general settings, our mechanism preserves the SW guarantee and continues to achieve the best possible guarantee even under polymatroid constraints. However, for the LW guarantee, there is a gap between our guarantee in Theorem~\ref{efficiency-1st} and the upper bound mentioned above. 
Since a mechanism proposed by \citet{LX2015} for one-sided markets achieves a better LW guarantee than clinching auctions (although it is currently not applicable to polymatroidal settings), a similar mechanism may also exist in two-sided markets. Identifying such a mechanism is a promising direction for future research. Even if one is found, we believe that our mechanism will remain valuable due to its broad applicability.

Improving the upper bound is another intriguing and challenging direction. Tackling this problem will require larger instances, as our mechanism already achieves the best possible in bilateral trade.
Previous results on upper bounds and impossibility theorems in two-sided markets are largely based on bilateral trade, overcoming certain barriers will be essential. Specifically, it may be necessary to analyze mechanisms where the transaction price depends not only on the sample values but also on many other factors such as budgets. Closing this gap by advancing in one of the above areas remains an important open problem.

Another direction concerns the budget balance property. While our mechanism satisfies WBB, the strongest form—direct-trade strong budget balance (DSBB)—requires that monetary transfers occur only through direct trades between participants, as introduced by \citet{BGKLR2020}. Exploring whether a mechanism can simultaneously satisfy DSIC, IR, and DSBB while achieving a constant approximation to the optimal LW is a promising avenue. Developing single-sample mechanisms that meet these criteria remains an important challenge for future research.

\section{Proof of Theorem \ref{payment}}
Following the outline in Section 4.2, we prove Theorem \ref{payment}, which is our main technical contribution. We focus on invariant properties that hold for any iteration of Algorithm \ref{HS_PCA}. Notably, such properties as those shown below have not been established, even in special cases of our setting.
\subsection{Simpler Formula for the Function $f_{x,d}$}	
	Let $(w,p,r)$ and $d$ be an allocation and demands in an iteration of Algorithm~\ref{HS_PCA}. Also, for the transaction vector $w$, define $x:=(x_i)_{i\in N}$, where $x_i=w(E_i)\ (i\in N)$. Our first result is~on the fundamental property of the function $f_{x,d}$. Specifically, we find that the formula (\ref{remnant}) in Theorem~\ref{clinch_sec3} can be simplified by using the clinching condition of ``not affecting other buyers.''  This result might clearly illustrate the key feature of the clinching framework.
	
	\begin{theorem}
	\label{invariant}
	Throughout the execution of Algorithm \ref{HS_PCA}, for each $S\subseteq N$, it holds 
	\begin{equation}
	\label{remaining supply}
	f_{x,d}(S)=\min_{S'\subseteq S}\{f(S')-x(S')+d(S\setminus S')\}.
	\end{equation}
	\end{theorem}
	This property is useful for analyzing the total number of goods allocated to buyers and their payments.
	In particular, we use this theorem to show Proposition \ref{remaining_goods}. 
	Now we prove Theorem~\ref{invariant} using mathematical induction. To simplify the notation, we define $\delta:=(\delta_i)_{i\in N}$ for a transaction vector $\xi_i$ of buyer $i$ in Algorithm~\ref{HS_clinch}, where $\delta_i=\xi_i(E_i)\ (i\in N)$. Then, the following holds:
	
	\begin{lemma}
	\label{clinch_amount_XY}
	For any $S,T\subseteq N$ with $S\subset T$, it holds
	$\delta(T\setminus S) \leq f_{x,d}(T)-f_{x,d}(S)$ in the iteration.
	\end{lemma}
	\begin{proof}
	Suppose that $T\setminus S=\{\ell_1, \ell_2, \ldots, \ell_q\}$ for some positive integer $q$.
	By Theorem~\ref{clinch_sec3} (ii), 
	it holds  
	$\delta_\ell=f_{x,d}(N)-f_{x,d}(N\setminus \ell)$ for each $\ell\in T\setminus S$.
	Then, we have
	\begin{align*}
	\delta(T\setminus S)&= \sum_{\ell\in T\setminus S}\left(f_{x,d}(N)-f_{x,d}(N\setminus \ell)\right)\\
	&\leq f_{x,d}(T)-f_{x,d}(T\setminus \ell_1)+f_{x,d}(T\setminus \ell_1)-\cdots +f_{x,d}(S\cup \ell_q)-f_{x,d}(S)
	=f_{x,d}(T)-f_{x,d}(S),
	\end{align*}
	where the inequality holds by the submodularity of $f_{x,d}$.
	\end{proof}
	
	\begin{proof}[Proof of Theorem~\ref{invariant}]	
	At the beginning of Algorithm~\ref{HS_PCA}, by $x_i=0$ for each $i\in N$ and the monotonicity of $f$, 
	we have $\min_{S'\supseteq S}\{f(S')-x(S')\}=\min_{S'\supseteq S} \{f(S')\}=f(S)$ for each $S\subseteq N$.
	We show that if (\ref{remaining supply}) holds for any 
	$S\subseteq N$ before the execution of a step, 
	it also holds after that.
	
	(i) Buyers clinch $\delta$ amount of goods in line 4: 
	By the inductive assumption, there exists $S'\subseteq S$ such that $f_{x,d}(S)=f(S')-x(S')+d(S\setminus S')$ for each $S\subseteq N$ just before line~4.
	When each buyer $i\in S$ clinches $\delta_i$,
	this increases $x_i$ by $\delta_i$ and decreases $d_i$ by $\delta_i$.
	Let $\tilde{x}$ and $\tilde{d}$ be the allocation and demands just after line~4 in the iteration.
	Whether these buyers belongs to $S'$ or $S\setminus S'$, it~holds 
	\begin{equation}
	\label{invariant_S}
	\min_{S'\subseteq S}\{f(S')-\tilde{x}(S')+\tilde{d}(S\setminus S')\}=
	\min_{S'\subseteq S}\{f(S')-x(S')+d(S\setminus S')\}-\delta(S)
	=f_{x,d}(S)-\delta(S).
	\end{equation}
	Suppose to the contrary that (\ref{remaining supply}) does not hold just after line 4.
	Then, we have 
	\begin{equation}
	\label{middle2}
	f_{\tilde{x},\tilde{d}}(S)<\min_{S'\subseteq S}\{f(S')-\tilde{x}(S')+\tilde{d}(S\setminus S')\}=f_{x,d}(S)-\delta(S).
	\end{equation}
	Now we use the following claim, where the proof is given in Appendix C.
	\begin{claim}
	\label{sets}
	In (\ref{remnant}), if 
	$f_{x,d}(S)<\min_{S'\subseteq S}\{f(S')-x(S')+d(S\setminus S')\}$ for some $S\subseteq N$, 
	then there exist $S^*\subseteq S$ and $S^{\dag} \supset S^*$ such that 
	$S^{\dag}\setminus S\neq \emptyset$, $(S^\dag\setminus S^*)\cap S=\emptyset$, and 
	$f_{x,d}(S)=f(S^\dag)-x(S^\dag)+d(S\setminus S^*)$.
	\end{claim}
	Define $T:=S^\dag\cup (S\setminus S^*)$. 
	By $T\subseteq N$, (\ref{invariant_S}) also holds for $T$ and thus we have
	\[
	f_{\tilde{x},\tilde{d}}(S)=f(S^\dag)-\tilde{x}(S^\dag)+\tilde{d}(S\setminus S^*)\geq 
	\min_{T'\subseteq T}\{f(T')-\tilde{x}(T')+\tilde{d}(T\setminus T')\}=f_{x,d}(T)-\delta(T).
	\]
	By (\ref{middle2}), we have
	$f_{x,d}(S)-\delta(S)>f_{x,d}(T)-\delta(T)$.
	By $S\subset (S^\dag\setminus S^*)\cup S=S^\dag\cup (S\setminus S^*)=T$, we have $\delta(T\setminus S)>f_{x,d}(T)-f_{x,d}(S)$, which contradicts Lemma~\ref{clinch_amount_XY}.
	Thus, (\ref{remaining supply}) also holds just after~line~4.

	(ii) The price update of buyer $l$ in line 5: 
	Since $x$ and $d$ are unchanged, (\ref{remaining supply}) still holds.

	(iii) The demand update of buyer $l$ in line 6: 
	Let $d'$ be the demands of buyers after the update. 
	Then, it holds $d'_l<d_l$ and $d'_i=d_i$ for each $i\in N\setminus l$.
	It suffices to consider the case of $l\in S$.
	If $l\in S'$ in (\ref{remnant}) after the demand update, by $d_l> d'_l$, 
	it also holds $l\in S'$ before the update and thus $f_{x,d'}(S)=f_{x,d}(S)$. 
	If $l\in S\setminus S'$ in (\ref{remnant}) after the demand update, it holds $f_{x,d'}(S)=f_{x,d}(S\setminus l)+d'_l$.
	Therefore, we have 
	$f_{x,d'}(S)=\min\{f_{x,d}(S), f_{x,d}(S\setminus l)+d'_l\}= \min_{S'\subseteq S}\{f(S')-x(S')+d'(S\setminus S')\}$, 
	which means that (\ref{remaining supply}) also holds just after line 6.
	
	Thus, if (\ref{remaining supply}) holds for any $S\subseteq N$ before the execution of a step, 
	it also holds after the~step.
	\end{proof}

\subsection{The Number of Remaining Goods}	
	Now we provide a lower bound on the number of remaining goods.
	Let $X:=\{i\in N\colon d_i>0\}$ denote the set of \textit{active} buyers, i.e., buyers with positive demands. 
	Using Theorem~\ref{invariant}, we now show that in any iteration of the mechanism, the remaining goods for active buyers $S$ are at least the difference between $x(S)$ and $x^{*}(S)$.
	
	\begin{proposition}
	\label{remaining_goods}
	Throughout the execution of Algorithm \ref{HS_PCA}, the following holds: 
	\begin{itemize}
	\item[(i)] It holds $f_{x,d}(N)=f(N)-x(N)$.
	\item[(ii)] It holds $f_{x,d}(X)=f(N)-x(X)-x^{\rm f}(N\setminus X)$. In particular, 
			$f_{x,d}(X)$ is unchanged by the demand update in line 6 of Algorithm \ref{HS_PCA}.
	\item[(iii)] For any $S\subseteq X$, it holds $f_{x,d}(S)\geq x^*(S)-x(S)$.
	\end{itemize}
	\end{proposition}

	\begin{proof}
	(i) This property is directly obtained by the combination of 
	Lemma 4.17 of \citet{GMP2014} and Theorem 7 of~\citet{HS2022}.
	
	(ii) For $i\in N\setminus X$, it holds $d_i=0$.
	Then, by Theorem~\ref{invariant}, it holds 
	\[
	f_{x,d}(X)=\min_{S\subseteq X}\{f(S)-x(S)+d(X\setminus S)\}
	=\min_{S\subseteq X}\{f(S)-x(S)+d(X\setminus S)\}+d(N\setminus X)\geq f_{x,d}(N).
	\]
	By the monotonicity of $f_{x,d}$, we have $f_{x,d}(X)=f_{x,d}(N)$. 
	Also, by (i), we have
	$f_{x,d}(X)=f_{x,d}(N)=f(N)-x(N)=f(N)-x(X)-x^{\rm f}(N\setminus X)$.
	The second statement is trivial from the first statement.
	
	(iii) For any $S\subseteq X$, let $S'\subseteq S$ be a minimizer of $f_{x,d}(S)$, i.e.,  
	$f_{x,d}(S)=f(S')-x(S')+d(S\setminus S')$ by Theorem~\ref{invariant}. 
	By $x^*(S')\leq f(S')$ (Proposition \ref{optimal}), we have the following:
	\begin{align*}
	f_{x,d}(S)-x^*(S)+x(S)&=f(S')-x^*(S')+x(S\setminus S')+d(S\setminus S')-x^*(S\setminus S')\\
	&\geq x(S\setminus S')+d(S\setminus S')-x^*(S\setminus S').
	\end{align*} 
	
	Let $c_i$ denote the current price of buyer $i$. 
	Suppose that $c_i>0$ for all $i\in S\setminus S'$.
	In the mechanism, buyer $i$ clinches the goods at the price equal to or less than $c_i$.
	Then, we have $x_i\geq p_i/c_i$ for each $i\in N$.
	Moreover, since buyer $i\in S\setminus S'$ is still active, we have $v_i\geq c_i$, and thus 
	$x^*_i\leq B_i/v_i\leq B_i/c_i$ by Proposition \ref{optimal}.
	Therefore, we have 
	\begin{align*}
	f_{x,d}(S)-x^*(S)+x(S)\geq \sum_{i\in S\setminus S'}(x_i+d_i-x^*_i)&\geq
	 \sum_{i\in S\setminus S'}\left(p_i/c_i+(B_i-p_i)/c_i-B_i/c_i\right)=0.
	\end{align*} 
	Suppose that $c_i=0$ for some $i\in S\setminus S'$. By $d_i=\infty$, 
	we also have $f_{x,d}(S)-x^*(S)+x(S)>0$.
	\end{proof}

Since all goods are sold by Theorem \ref{properties_sec3} (ii), multiplying the current price by the number of remaining goods provides a lower bound on future payments of buyers. However, all payments of virtual buyers are canceled at the end of the auction. A more detailed analysis is therefore required.

\subsection{A Lower Bound on Future Payments}
To provide a lower bound on future payments of active \textit{non-virtual} buyers, 
we define a set $Y:=\{i\in X\colon x^{\rm f}_i\leq x^{*}_i\}$. By Proposition \ref{x>xopt}, the set $Y$ contains all active virtual buyers. The following inequality provides the desired lower bound for active non-virtual buyers in $X \setminus Y$:
	
	\begin{proposition}
	\label{payment2}	
	It holds throughout the auction that 
	\begin{equation}
	\label{payment3}
	\sum_{i\in X\setminus Y}(p^{\rm f}_i-p_i)\geq\sum_{i\in Y}(v_i-\varepsilon)(x_{i}^*-x^{\rm f}_{i})+\tilde{c} (f_{x,d}(X)-x^*(Y)+x(Y)),
	\end{equation}
	where $\tilde{c}:=\min_{i\in X\setminus V}c_i$. 
	\end{proposition}
	
	For the proof, we show that if (\ref{payment3}) holds 
	at the end of an iteration, it also holds at the beginning of the iteration. 
	To the best of our knowledge, this is the first study to use a type of \textit{backward} mathematical induction  
	for the polyhedral clinching auction.

	\begin{proof}
	At the end of the auction, we have (\ref{payment3}) 
	by $X=Y=\emptyset$ and $x^{\rm f}(N)=f(N)$ from Theorem~\ref{properties_sec3}~(ii).
	By the following case-by-case analysis, 
	we show that if (\ref{payment3}) holds at the end of an iteration, 
	then it holds at the beginning of the iteration.

	(i) The price update in line 5: 
	The left-hand side and the first term on the right-hand side of (\ref{payment3}) are unchanged.
	Then, by Proposition \ref{remaining_goods} (iii), it holds 
	$f_{x,d}(X)-x^*(Y)+x(Y)\geq f_{x,d}(Y)-x^*(Y)+x(Y) \geq 0$, 
	where the first inequality holds by $Y\subseteq X$ and the monotonicity of $f_{x,d}$. 
	This guarantees that the second term on the right-hand side is non-decreasing.
	This means that the right-hand side is non-decreasing by the price increase of buyer $l$.
	Therefore, if (\ref{payment3}) holds after the price update, it also holds before the update.

	(ii) The update of the demand in line 6: By Proposition \ref{remaining_goods} (ii), we can see that $f_{x,d}(X)$ is independent of demands.
	Then, when the demand $d_l$ of buyer $l$ is slightly decreased by the price increase, 
	both sides of (\ref{payment3}) are unchanged since both sides are independent of demands.
	Therefore, it suffices to consider the case where buyer $l$ drops out of the auction by $c_l=v_l$.
	Suppose that $l\in X\setminus Y$ just before line 6 in the iteration.
	The left-hand side of (\ref{payment3}) is unchanged 
	by $p_l=p^{\rm f}_l$ before the demand update.
	On the right-hand side, the first term is unchanged by $l\notin Y$.
	The second term is also unchanged by Proposition \ref{remaining_goods} (ii) because $l$ is just moved from 
	$X\setminus Y$ to $N\setminus X$ and it holds $x_l=x^{\rm f}_l$.
	From the above, both sides of (\ref{payment3}) are unchanged. 
	Suppose that $l\in Y$ just before line 6 in the iteration.
	The left-hand side of (\ref{payment3}) is trivially unchanged by $l\notin X\setminus Y$.
	On the right-hand side, the first term is decreased by $(v_l-\varepsilon)(x^*_i-x^{\rm f}_i)$ and 
	the second term is increased by $\tilde{c}(x^*_i-x^{\rm f}_i)$ because $l$ is added to $N\setminus X$.
	By the demand update, the right-hand side is increased by 
	$\left(\tilde{c}-(v_l-\varepsilon)\right)(x^*_i-x^{\rm f}_i)$.
	Since $\tilde{c}\in \{v_l, v_l-\varepsilon\}$, the right-hand side is not decreased by the demand update.
	This means that if (\ref{payment3}) holds after the demand update, it also holds before the update.
	
	(iii) Clinching step in line 4: 
	We consider the case where buyer $i\in N$ clinches $\xi_i(E_i)$ amount of goods.
	Suppose that $i\in Y$ just before line 4 in the iteration.
	If it holds $p^{\rm f}_i=B_i$, then we have $x^{\rm f}_i> p^{\rm f}_i/v_i=B_i/v_i \geq x^*_i$, 
	which contradicts $i\in Y$.
	Therefore, we have $p^{\rm f}_i<B_i$.
	By Proposition~\ref{dropping}, $i$ does not drop out of the auction after the clinching step.
	Then, the left-hand side of (\ref{payment3}) and the first term on the right-hand side are unchanged.
	The second term on the right-hand side is also unchanged 
	by $f_{x,d}(X)-x^*(Y)+x(Y)=f(N)-x(X\setminus Y)-x^*(Y)-x^{\rm f}(N\setminus X)$ 
	(Proposition~\ref{remaining_goods}~(ii)).
	Suppose that $i\in X\setminus Y$ just before line 4 in the iteration.
	The left-hand side of (\ref{payment3}) is decreased by $c_i \xi_i(E_i)$.
	For the right-hand side, the first term is unchanged by $l\notin Y$ and 
	the second term is decreased by $\tilde{c}\xi_i(E_i)$ 
	regardless of whether $i$ belongs to $X\setminus Y$ or $N\setminus X$ after the clinching step.
	Therefore, if (\ref{payment3}) holds after the clinching step, it holds before that.
	\end{proof}
	
\subsection{A Lower Bound on Total Payment}
	Using Proposition \ref{payment2}, we prove Theorem \ref{payment} 
	by considering the initial step of the auction.
	In the proof, we use the following lemma:
	\begin{lemma}
	\label{deletevarepsilon}
	If $\varepsilon\leq v_{min}^2/(v_{max}-v_{min})$ 
	for each $i\in N$ with $x^{\rm f}_i>x^*_i$, it holds $(v_{min}+\varepsilon)x^*_i-\varepsilon x^{\rm f}_i\geq 0$.
	\end{lemma}
	\begin{proof}
	For $i\in N$ with $x^{\rm f}_i>x^*_i$, by Proposition \ref{x>xopt}, it holds $x^*_i=B_i/v_i$ 
	and $i$ is a non-virtual buyer.
	Thus, by Theorem \ref{clinch_sec3} (iii), 
	the minimum transaction price of buyer $i$ is at least $v_{min}$.
	Combining this with budget feasibility, we have $x^{\rm f}_i\leq p^{\rm f}_i/v_{min}\leq B_i/v_{min}$.
	Therefore, by $v_i\leq v_{max}$, it holds  
	\begin{align*}
	\left(v_{min}+\varepsilon\right) x^*_i-\varepsilon x^{\rm f}_i \geq B_i \left(\frac{v_{min}+\varepsilon}{v_{i}}-\frac{\varepsilon}{v_{min}}\right)\geq B_i \left(\frac{v_{min}+\varepsilon}{v_{max}}-\frac{\varepsilon}{v_{min}}\right).
	\end{align*}
	It suffices to show that the right-hand side is nonnegative.
	By $\varepsilon\leq v_{min}^2/(v_{max}-v_{min})$, we have
	\[
	B_i \left(\frac{v_{min}+\varepsilon}{v_{max}}-\frac{\varepsilon}{v_{min}}\right)=
	B_i\left(\frac{v_{min}}{v_{max}}-\varepsilon\left(\frac{1}{v_{min}}-\frac{1}{v_{max}}\right)\right)
	\geq B_i\left(\frac{v_{min}}{v_{max}}-\frac{v_{min}}{v_{max}}\right)=0.
	\]
	\end{proof}	

	\begin{proof}[Proof of Theorem \ref{payment}]
	Consider the iteration where all the price clocks of non-virtual buyers just reach $\min_{i\in V}v_i$. 
	At the moment, all virtual buyers are still active by $c_{n+j}<v_{n+j}=\rho_j$ for each $j\in M$. 
	Then, by Theorem~\ref{clinch_sec3} (iii), 
	it holds $x_i=p_i=0$ for each $i\in N\setminus V$.
	Combining this with the second statement of Proposition \ref{x>xopt}, 
	we have $V\subseteq Y\subseteq X\subseteq N$. Therefore, we have 
	\begin{equation}
	\label{relations}
	x(V)=x(Y)=x(X)=x(N)\ {\rm and}\ x^{\rm f}_i=x_i=0\ \ (i\in N\setminus X),
	\end{equation}
	where the last equality holds by the definition of $X$.
	By the last statement of Proposition~\ref{optimal}, we have  $x_i^{*}=x^{\rm f}_i=0$ for each $i\in N\setminus X$.
	This implies that in the iteration, we have 
	\begin{equation}
	\label{sets2}
	N\setminus Y=X\setminus Y=\{i\in N\colon x^{\rm f}_i> x^{*}_i\}\ {\rm and}\ Y=\{i\in N\colon x^{\rm f}_i< x^{*}_i\}\cup 
	\{i\in X\colon x^{\rm f}_i=x^{*}_i\}.
	\end{equation}
	Then, by Proposition \ref{remaining_goods} (ii) and $x^*(N)=f(N)$ (Proposition \ref{optimal}), we also have 
	\begin{align*}
	f_{x,d}(X)-x^*(Y)+x(Y)= f(N)-x^*(Y) = \sum_{i\in N; x^{\rm f}>x^*_i}x^*_i,
	\end{align*}
	where the first equality holds by $x(X)=x(Y)$ and $x^{\rm f}(N\setminus X)=x(N\setminus X)=0$ from (\ref{relations}).

	By Proposition \ref{payment2}, (\ref{payment3}) holds throughout the auction.
	Substituting (\ref{sets2}) with (\ref{payment3}), we have 
	\begin{align}
	\label{eq1}
	\sum_{i\in N; x^{\rm f}_i>x^{*}_i} p^{\rm f}_i&=\sum_{i\in X\setminus Y} p^{\rm f}_i \geq 
	\sum_{i\in Y}(v_i-\varepsilon)(x_{i}^*-x^{\rm f}_{i})+ \tilde{c} (f_{x,d}(X)-x^*(Y)+x(Y)) \nonumber\\
	&\geq \sum_{i\in N; x^{\rm f}_i\leq x^{*}_i}(v_i-\varepsilon)(x_{i}^*-x^{\rm f}_{i})+v_{min} \sum_{i\in N; x^{\rm f}>x^*_i}x^*_i,
	\end{align}
	where the second inequality holds by $\tilde{c}=\min_{i\in V}v_i \geq v_{min}$.
	Then, by Lemma \ref{deletevarepsilon}, we have 
	\[
	-\varepsilon\sum_{i\in N; x^{\rm f}_i\leq x^*_i}(x_{i}^*-x^{\rm f}_{i})+v_{min}
	\sum_{i\in N; x^{\rm f}>x^*_i}x^*_i=\sum_{i\in N; x^{\rm f}_i> x^*_i} \left((v_{min}+\varepsilon)x^*_i-\varepsilon x^{\rm f}_i\right)\geq 0,
	\]
	where the first equality holds by $x^*(N)=x^{\rm f}(N)=f(N)$ from Theorem \ref{properties_sec3} (ii) and Proposition~\ref{optimal}.
	Therefore, by (\ref{eq1}), we have $\sum_{i\in N; x^{\rm f}_i> x^*_i}p^{\rm f}_i \geq\sum_{i\in N; x^{\rm f}_i\leq x^*_i}v_i(x_{i}^*-x^{\rm f}_{i})$.
\end{proof}

\appendix
\section{LW Guarantee for Our Mechanism in Bilateral Trade}
We show that in the case where $n=m=1$ and $s_1=1$, 
our single-sample mechanism achieves a 1/2-approximation to the optimal LW.
Let $\rho_a$ and $\rho_b$ denote the seller's valuation and sample value, respectively.
We change these values and fix the distribution and the information on the buyer.

If $\rho_a=\rho_b$, this reduces to the case of a truthful seller and 
by Theorem~\ref{LWofHS}, our mechanism achieves a 1/2-approximation.
Then, we can assume $\rho_a>\rho_b$. 
Let ${\rm LW}^{\mathbf  M_{\rm sample}(\rho, \rho_s)}$ be the LW value of Algorithm \ref{First} 
when the valuation and the sample value are $\rho$ and $\rho_s$, respectively, 
and ${\rm LW}^{{\rm OPT}(\rho)}$ be the optimal LW value when the valuation is $\rho$.
By Lemma~\ref{Approx}, it suffices to show:
\[
{\rm LW}^{\mathbf M_{\rm sample}(\rho_a, \rho_b)}+{\rm LW}^{\mathbf M_{\rm sample}(\rho_b, \rho_a)}\geq {\rm LW}^{{\rm OPT}(\rho_a)}\geq\frac{1}{2}({\rm LW}^{{\rm OPT}(\rho_a)}+ {\rm LW}^{{\rm OPT}(\rho_b)}).
\]

The second inequality holds trivially by $\rho_a>\rho_b$. Now we show the first inequality. 
When $\rho_a$ is the valuation and $\rho_b$ is the sample value, in Algorithm \ref{First}, 
the seller does not participate in the auction, and thus we have 
${\rm LW}^{\mathbf M_{\rm sample}(\rho^a, \rho^b)}=\rho_a$. 
When $\rho_b$ is the valuation and $\rho_a$ is the sample value,
we perform the case-by-case analysis according to whether the good is divisible or indivisible.
\paragraph{Divisible Case.}
Suppose that $\rho_a<v_1$. In Algorithm \ref{HS_PCA}, by Proposition \ref{dropping}, 
the buyer exhausts her budget if $B_1\leq\rho_a<v_1$ and 
all the good is allocated to the buyer otherwise.
Since the allocation for the buyer is directly used in Algorithm \ref{First},  
we have ${\rm LW}^{\mathbf M_{\rm sample}(\rho_a, \rho_b)}=\min(v_1, B_1)$ and thus 
\[
{\rm LW}^{\mathbf M_{\rm sample}(\rho_a, \rho_b)}+{\rm LW}^{\mathbf M_{\rm sample}(\rho_b, \rho_a)}=
\min(v_1, B_1)+\rho_a\geq  {\rm LW}^{{\rm OPT}(\rho_a)}.
\]
Suppose that $v_1\leq \rho_a$. Then, the optimal LW value is achieved when all the good is allocated to the seller.
Therefore, we have
${\rm LW}^{\mathbf M_{\rm sample}(\rho_a, \rho_b)}=\rho_a={\rm LW}^{{\rm OPT}(\rho_a)}$.

\paragraph{Indivisible Case.}
Suppose that $\rho_a<\min(v_1,B_1)$. In Algorithm \ref{Indivisible3}, all the good is allocated to the buyer.
As in the divisible case, we have ${\rm LW}^{\mathbf M_{\rm sample}(\rho_a, \rho_b)}=\min(v_1, B_1)$ and thus 
\[
{\rm LW}^{\mathbf M_{\rm sample}(\rho_a, \rho_b)}+{\rm LW}^{\mathbf M_{\rm sample}(\rho_b, \rho_a)}=
\min(v_1, B_1)+\rho_a\geq  {\rm LW}^{{\rm OPT}(\rho_a)}.
\]
Suppose that $\min(v_1,B_1)\leq \rho_a$. As in the divisible case, we have~${\rm LW}^{\mathbf M_{\rm sample}(\rho_a, \rho_b)}={\rm LW}^{{\rm OPT}(\rho_a)}$.

\section{Polymatroids}
Since we use polymatroid theory in the appendix, we begin with a brief overview of the theory. For more information on the basics of polymatroids, see \citep{F2005,S2003}.
\subsection{Polymatroids}
A polymatroid $P:=P(f)$ on $N$ associated with the monotone submodular function $f:2^N\to\mathbf R_+$ 
is a polytope defined by $P(f):=\{x\in \mathbf R^N_+: x(S)\leq f(S)\ (S\subseteq N)\}$.
We often denote $P(f)$ by $P$. Then, it is known that there is a one-to-one correspondence between 
the polymatroid $P$ and the monotone submodular function~$f$:
\begin{lemma}[e.g., \citet{S2003}]
\label{polymatroid_equal}
Two polymatroids are equal if and only if the monotone submodular functions corresponding to the polymatroids are equal.
\end{lemma}
Polymatroids are closed under various operations. Of these operations, we use the following: 
\begin{itemize}

\item \textbf{Contraction:}
For $x \in P$, define a polytope $P^x:=\{y \colon y \in P, y \geq x\}$,
where each component is more than that of~$x$.
Then, $P^x +(-x)$ is a polymatroid, where + denotes the Minkowski sum.
The monotone submodular function $f^x: 2^{N} \to \mathbf{R}_+$ that defines $P^x+ (-x)$ is expressed~as 
\[
f^x(S) := \min_{S' \supseteq S} \{f(S') -x(S' \setminus S)\}\quad (S\subseteq N).
\]
\item \textbf{Reduction:}
For $x \in \mathbf{R}^{N}$, define a polytope $P_x:=\{y \colon y \in P, y \leq x\}$, 
where each component is less than that of~$x$.
Then, $P_x$ is a polymatroid and the monotone submodular function 
$f_x: 2^{N} \to \mathbf{R}_+$ that defines $P_x$ is expressed~as
\[
f_x(S) := \min_{S' \subseteq S}\{f(S') +x(S \setminus S')\}\quad (S\subseteq N).
\]

\item \textbf{Induction:}
Let $(N, M, E)$ be a bipartite graph and $P$ be a polymatroid on $M$ 
associated with a monotone submodular function $f:2^M\to \mathbf R_{+}$.
Now define a function $g:2^N\to \mathbf R_{+}$ by $g(S):=f(M_S)\ \, (S\subseteq N)$.
Then, $g$ is a monotone submodular function and the polymatroid $P':=P'(g)\subseteq R^N_{+}$ is expressed~as 
\[
P':=\{x\in \mathbf R_+^{N}: \exists w\in P,\ x_i=w(E_i)\  (i\in N)\}.
\]
\end{itemize}

Subsequently, for a polymatroid $P$ on $N:=\{1,2,\ldots,n\}$ 
associated with a monotone submodular function $f: 2^N\to\mathbf R_+$, 
consider the following optimization problem:
\begin{align*}
{\rm maximize}\quad \sum_{i \in N}\omega_i x_i \qquad
{\rm s.t.}\quad x \in P,
\end{align*}
where $\omega:=(\omega_i)_{i\in N}\in \mathbf R^N_+$ denotes the weights of the elements.
Then, \citet{E1970} showed that this problem can be  solved efficiently by the following greedy procedure:
\begin{itemize}
\item[1.] Arrange the elements in $N$ so that $\omega_1 \geq \omega_2 \geq \cdots \geq \omega_n$.
\item[2.] For $i=1,2,\ldots, n$, define $X_i:=\{1,2,\ldots,i\}$ and $x_i:= f(X_i) - f(X_{i-1})$.
\item[3.] Output the value of the objective function for $x:=(x_i)_{i\in N}$.
\end{itemize}
The optimal solution $x$ is on the base polytope $B(f) := \{y \colon y \in P, y(N) = f(N)\}$.
This also implies that a point on $B(f)$ can be computed in polynomial time, 
if the value oracle of $f$ is given.

\subsection{Polymatroid Flow Problems}
A polymatroidal network consists of~a directed network $(V, E)$ with a source $s$ and a sink $t$, 
where each node $v\in V$ has associated polymatroids $P^{+}_v$ and $P^{-}_v$. Let $\delta_v^+$ and $\delta_v^-$ denote the sets of edges in $E$ leaving and entering $v$, respectively. 
A flow $\varphi: E \to \mathbf{R}+$ in this network is defined as a function satisfying the following~properties:
\begin{align*}
\sum_{e \in \delta^+_v}\varphi(e)= \sum_{e \in \delta^-_v}\varphi(e) \quad (v \in V\setminus\{s,t\}),\ \ 
\varphi|_{\delta^+_v} \in P_v^+, \varphi|_{\delta^-_v} \in P_v^-.
\end{align*}
If the network has an edge capacity $c: E \to \mathbf{R}_+$, 
the polymatroid that represents the capacity constraint is given by the submodular function
$F \mapsto \sum_{e \in F}c(e) \quad (F \subseteq \delta_v^+\ ({\rm or} \ \delta_v^-))$.
The flow-value of a flow $\varphi$ is expressed as
\[
\displaystyle \sum_{e \in \delta^+_s}\varphi(e) - \sum_{e \in \delta^-_s}\varphi(e) ( = \sum_{e \in \delta^-_t}\varphi(e) - \sum_{e \in \delta^+_t}\varphi(e)).
\]

Now we introduce the max-flow min-cut theorem in polymatroidal networks. A cut $(S,T)$ of sets of nodes such that $s \in S$ and $t \in T$, and a partition $(A,B)$ of the set of edges going from $S$ to $T$, is called a arc-partitioned cut 
$(S,T,A,B)$. Note that $S \cup T = V$. 
The capacity of this arc-partitioned cut is expressed as
$\displaystyle \sum_{v \in T} f^-_v(\delta^-_v\cap A) + \sum_{v \in S}f^+_v(\delta^+_v\cap B)$, 
where $f^+_v$ and $f^-_v$ are the monotone submodular functions for $P^+_v$ and $P^-_v$, respectively. 
\citet{LM1982} showed that 
the maximum value of a flow in a polymatroid network is equal to 
the minimum capacity of an arc-partitioned cut. They also proposed an algorithm~to compute the maximum flow value in polynomial time, given the value oracles of the submodular functions.

\section{Omitted Proofs}
Define $P:=\{w\in \mathbf R^{E}_+ \colon w(E_j)\leq s_j\ (j\in M)\}$.
Consider a polymatroid $\tilde{P}\subseteq \mathbf R^{N}_+$ 
defined by the monotone submodular function $f$. 
Since $\tilde{P}$ is also obtained by the induction of $P$, then it holds 
	\begin{equation}
	\label{tildeP}
	\tilde{P}:
	=\{x\in \mathbf R_{+}^{N}\colon 
	 \exists w\in P,\ 
	 x_i=\sum_{ij\in E_i}w_{ij} \quad (i\in N) \}.
	\end{equation}
\subsection{Proof of Proposition \ref{optimal}}
The LW maximization problem can be represented by:
	\begin{align}
	\label{optimization_sec3}
	{\text{maximize}}\ \ \  \sum_{i\in N}\min(v_i x_i,B_i)
	\quad \text{subject to}\ \ \  x\in \tilde{P} \nonumber.
	\end{align}
	If $x_i\geq B_i/v_i$ for some $i\in N$, the value $\min(v_i x_i,B_i)$ does not change as $x_i$ increases.  
	This implies that we can restrict $x$ by $x_i\leq B_i/v_i$ for each $i \in N$ 
	and reformulate the problem by 
	\begin{equation}
	\label{linear_opt}
	{\text{maximize}}\ \ \  \sum_{i\in N}v_i x_i
	\quad \text{subject to}\ \ \  x\in \tilde{P}_{d^*},
	\end{equation}
	where $d^{*}\in \mathbf R^N_{+}$ is defined by $d^{*}_i:=B_i/v_i$ for each $i\in N$ and 
	$\tilde{P}_{d^*}:=\{y\in \tilde{P}\colon y_i\leq d^*_i\ (i\in N)\}$ is a polymatroid 
	obtained by the reduction of the polymatroid $\tilde{P}$ by the vector $d^*$.
	Then, the monotone submodular function $f_{d^*}$ associated with $\tilde{P}_{d^*}$ is represented by 
	\begin{equation}
	\label{f_d}
	f_{d^*}(S)=\min_{T\subseteq S}\{f(S\setminus T)+d^*(T)\}.
	\end{equation}
	The problem (\ref{linear_opt}) can be viewed as a linear optimization on a polymatroid, 
	which can be solved by the greedy procedure.
	Then, we have $x^*(S)\leq f_{d^*}(S)\leq f(S)\ (S\subseteq N)$ and $x^*_i=f_{d^*}(H_i\cup i)-f_{d^*}(H_i)$. 
	Below, we show the remaining part using the following lemma and (\ref{f_d}):
\begin{lemma}[e.g. \citet{F2005}]
	\label{tightsets_union}
	Let $P$ be a polymatroid defined by a  monotone submodular function $f:2^{N}\to \mathbf R_{+}$.
	For a vector $x\in P$, if it holds $x(S)=f(S)$ and $x(T)=f(T)$ for some $S, T\subseteq N$,
	then it also holds $x(S\cap T)=f(S\cap T)$ and $x(S\cup T)=f(S\cup T)$.
	\end{lemma}
\begin{proof}[Proof of Proposition \ref{optimal}]
	Suppose that $f_{d^*}(H_i\cup i)=f(H)+d^*(H_i\setminus H)+d^*_i$ for $i\in N$ and some $H\subseteq H_i$.
	This means that $\min_{H'\subseteq H_i}\{f(H_i\setminus H')+d(H')\}=f(H)+d^*(H_i\setminus H)$.
	Then, it holds $f_{d^*}(H_i)=f(H)+d^*(H_i\setminus H)$.
	Therefore, we have $x^*_i=f_{d^*}(H_i\cup i)-f_{d^*}(H_i)=d^*_i$. 
	
	Suppose that $f_{d^*}(H_i\cup i)=f(H\cup i)+d^*(H_i\setminus H)$ for $i\in N$ and some $H\subseteq H_i$.
	This means 
	\[
	d^*(H_i\setminus H)=f_{d^*}(H_i\cup i)-f(H\cup i)\leq f_{d^*}(H_i\cup i)-f_{d^*}(H\cup i)\leq f_{d^*}(H_i)-f_{d^*}(H),
	\]
	where the first inequality holds by the definition of $f_{d^*}$, 
	and the second inequality holds by submodularity of $f_{d^*}$.
	By $x^{*}\in P_{d^*}$, it holds $x^{*}(H)\leq f_{d^*}(H)$ and $x^{*}_i\leq B_i/v_i=d^{*}_i$ for each $i\in N$.
	Using these inequalities, we have 
	\begin{align*}
	x^{*}(H_i)&=x^{*}(H)+x^{*}(H_i\setminus H)\leq f_{d^*}(H)+d^{*}(H_i\setminus H)\\
	&\leq f_{d^*}(H)+\left(f_{d^*}(H_i)-f_{d^*}(H)\right)=f_{d^{*}}(H_i)=x^{*}(H_i),
	\end{align*}
	where the last equality holds by the greedy procedure.
	This implies that all the above inequalities hold in equality.
	Then, we have $x^{*}(H)=f_{d^*}(H)$ and $x^{*}(H_i\setminus H)=d^{*}(H_i\setminus H)$, and thus 
	\[
	x^*_i=f_{d^*}(H_i\cup i)-f_{d^*}(H_i)=f(H\cup i)+d^*(H_i\setminus H)-x^{*}(H_i)
	=f(H\cup i)-x^{*}(H).
	\]
	In this case, $x^*_{i}$ is given as the minimum of $f(H\cup i)-x^{*}(H)$ with respect to $H\subseteq H_i$.
	
	Therefore, $x^{*}_i$ is recursively given by the following formula:
	\[
	 x^{*}_i=\min\left(B_i/v_i,{\min_{H\subseteq H_{i}}\{f(H\cup i)-x^{*}(H)}\}\right)\ (i\in N).
	\]
	
	Next, we show that $x^*(N)=f(N)$.
	Let $\{\eta_1,\eta_2,\ldots,\eta_m\}$ be the set of virtual buyers in descending order of valuations. 
	For each $\eta_k\in \{\eta_1,\eta_2,\ldots,\eta_m\}$, by $B_{\eta_k}=\infty$, we~have 
	\[
	x^{*}_{\eta_k}=\min_{H\subseteq H_{\eta_k}}\{f(H\cup\eta_k)-x^*_{\eta_k}(H)\}=f(H^*_{\eta_k})-x^*(H^*_{\eta_k}\setminus \eta_k)
	\]
	for some $H^*_{\eta_k}\subseteq H_{\eta_k}\cup \eta_k$ with $\eta_k\in H^*_{\eta_k}$. 
	This means that the sets $H^*_{\eta_1},\ H^*_{\eta_2},\ldots, H^*_{\eta_m}$ are all \textit{tight} sets.
	Then, we define $H^*$ as the union of all $H^*_{\eta_1},\ H^*_{\eta_2},\ldots, H^*_{\eta_m}$. 
	By Lemma \ref{tightsets_union}, $H^*$ is also a tight set.
	Therefore, by the monotonicity of $f$, we have 
	\begin{equation*}
	x^*(N)\geq x^*(H^*)=f(H^*)=f(H^*_{\eta_1}\cup\ldots\cup H^*_{\eta_m})\geq f(V)=\sum_{j\in M}s_j=f(N).
	\end{equation*}
	The opposite side holds trivially by the polymatroid constraint.
	Therefore, we have $x^*(N)=f(N)$.
	Moreover, by construction of the set $H^*$, 
	any non-virtual buyer $i$ with $v_i\leq \min_{j\in M}\rho_j$ is not included in the tight set $H^*$.
	Since $x^*(H^*)=f(N)$, we have $x^{*}_{i}=0$ for such buyer $i$. 
	\end{proof}

\subsection{Other Omitted Proofs}

\begin{proof}[Proof of Lemma \ref{relation_optimal}]
	In the following, we show the following inequality:
	\begin{align*}
	{\rm LW}^{{\rm OPT}|_{M_a}(\rho_a)}+
	{\rm LW}^{{\rm OPT}|_{M_b}(\rho_b)}
	\geq
	{\rm LW}^{{\rm OPT}(\rho^{\max})}
	\geq \frac{1}{2}({\rm LW}^{{\rm OPT}(\rho^a)}+{\rm LW}^{{\rm OPT}(\rho^b)}),
	\end{align*}
	where $\rho^{\max}:=\{\rho_j^{\max}\}_{j\in M}$ with $\rho^{\max}_j=\max(\rho^a_j, \rho^b_j)$ for each $j\in M$. 
	
	To show the first inequality,  
	let $\tilde{x}^{*}$ denote the LW optimal allocation that achieves ${\rm LW}^{{\rm OPT}(\rho^{\max})}$ 
	obtained by Proposition \ref{optimal}.
	Then, it holds $\min(v_i\tilde{x}^{*}_i, B_i)=v_i \tilde{x}^{*}_i$ for each $i\in N$.
	Thus, we have 
	\begin{align*}
	{\rm LW}^{{\rm OPT}(\rho^{\max})}&=\sum_{i\in N\setminus  V}v_i \tilde{x}^{*}_i+
	\sum_{j\in M_a}\rho^a_j \tilde{x}^{*}_{n+j}+
	\sum_{j\in M_b}\rho^b_j \tilde{x}^{*}_{n+j}-\sum_{j\in M: \rho^a_j=\rho^b_j}\rho^a_j \tilde{x}^{*}_{n+j}\\
	&\leq \sum_{i\in N\setminus  V}v_i \tilde{x}^{*}_i+
	\sum_{j\in M_a}\rho^a_j \tilde{x}^{*}_{n+j}+
	\sum_{j\in M_b}\rho^b_j \tilde{x}^{*}_{n+j}.
	\end{align*}
	
	By $\tilde{x}^{*}\in \tilde{P}$ (Proposition \ref{optimal}) and (\ref{tildeP}), 
	there exists $\tilde{w}^{*}\in P$ such that $\tilde{x}^{*}_i=\tilde{w}^{*}(E_i)\ (i\in N)$. Then, we have the following:
	\begin{align}
	\label{re_optimal}
	{\rm LW}^{{\rm OPT}(\rho^{\max})}&\leq \sum_{i\in N\setminus  V}v_i \tilde{x}^{*}_i+
	\sum_{j\in M_a}\rho^a_j \tilde{x}^{*}_{n+j}+
	\sum_{j\in M_b}\rho^b_j \tilde{x}^{*}_{n+j} \nonumber\\ 
	&\leq \sum_{j\in M_a}\left(\sum_{i\in N\setminus  V}v_i \tilde{w}^{*}_{ij}+\rho^a_j \tilde{w}^{*}(E_{n+j})\right)+
	\sum_{j\in M_b}\left(\sum_{i\in N\setminus  V}v_i \tilde{w}^{*}_{ij}+
	\rho^b_j \tilde{w}^{*}(E_{n+j})\right).
	\end{align}
	By $\tilde{w}^{*}\in P$, it holds $\tilde{w}^{*}(E_j)\leq s_j\ (j\in M)$.
	Then, this implies that, by the optimality, 
	the first (resp. second) term on the right-hand side in (\ref{re_optimal}) is bounded by ${\rm LW}^{{\rm OPT}|_{M_a}(\rho_a)}$
	(resp. ${\rm LW}^{{\rm OPT}|_{M_b}(\rho_b)}$).
	Therefore, we have ${\rm LW}^{{\rm OPT}|_{M_a}(\rho_a)}
	+{\rm LW}^{{\rm OPT}|_{M_b}(\rho_b)}\geq
	{\rm LW}^{{\rm OPT}(\rho^{\max})}$.

	Subsequently, we show the second inequality.
	Let $x^{*a}\in \tilde{P}$ be the allocation of goods 
	that achieves ${\rm LW}^{{\rm OPT}(\rho^a)}$.
	Then, we have
	\begin{align*}
	{\rm LW}^{{\rm OPT}(\rho^{\max})}&\geq \sum_{i\in N\setminus  V}v_i x^{*a}_i+
	\sum_{j\in M_a}\rho^a_j x^{*a}_{n+j}+
	\sum_{j\in M\setminus M_a}\rho^b_j x^{*a}_{n+j}\geq{\rm LW}^{{\rm OPT}(\rho^a)}, 
	\end{align*}
	where the first inequality holds by the optimality of ${\rm LW}^{{\rm OPT}(\rho^{\max})}$, and  
	the second inequality holds by $\rho^a_j\leq \rho^b_j$  for each $j\in M\setminus M_a$.
	Similarly, we have ${\rm LW}^{{\rm OPT}(\rho^{\max})}\geq {\rm LW}^{{\rm OPT}(\rho^b)}$. 
	Therefore, we have 
	${\rm LW}^{{\rm OPT}(\rho^{\max})}\geq \max({\rm LW}^{{\rm OPT}(\rho^a)},{\rm LW}^{{\rm OPT}(\rho^b)})\geq \frac{1}{2}({\rm LW}^{{\rm OPT}(\rho^a)}+{\rm LW}^{{\rm OPT}(\rho^b)}).$
\end{proof}

	\begin{proof}[Proof of the Claim in Theorem~\ref{invariant}]
	By (\ref{remnant}), there exist $S^*\subseteq S$ and $S^{\dag} \supseteq S^*$ such that 
	$f_{x,d}(S)=f(S^\dag)-x(S^\dag)+d(S\setminus S^*)$.
	Suppose that $S^{\dag}\subseteq S$. By $S^{\dag}\supseteq S^*$, it holds 
	\[
	\min_{S'\subseteq S}\{f(S')-x(S')+d(S\setminus S')\}\leq 
	f(S^\dag)-x(S^\dag)+d(S\setminus S^{\dag})\leq f(S^\dag)-x(S^\dag)+d(S\setminus S^*)=f_{x,d}(S), 
	\]
	which is a contradiction. Then, it holds $S^{\dag}\setminus S\neq \emptyset$. 
	Suppose that $(S^\dag\setminus S^*)\cap S\neq\emptyset$.
	Define $S^{**}:=((S^\dag\setminus S^*)\cap S)\cup S^*$. 
	By the definition of $S^{**}$, we have $(S^\dag\setminus S^{**})\cap S=\emptyset$.
	We have
	\[
	f(S^\dag)-x(S^\dag)+d(S\setminus S^*)
	\geq f(S^\dag)-x(S^\dag)+d(S\setminus S^{**}).
	\]
	Since it holds $S^{**}\subseteq S$ and $S^{\dag}\supseteq S^{**}$, 
	the claim holds when $S^{*}$ is replaced by $S^{**}$.
	\end{proof}

	\section*{Acknowledgement}
	We thank Hiroshi Hirai for careful reading and helpful comments.
	This work was supported by~Grant-in-Aid for JSPS Research Fellow Grant Number JP22J22831,~Grant-in-Aid~for~Challenging Research (Exploratory) Grant Number JP21K19759, and JST ERATO~Grant~Number~JPMJER2301.

\bibliography{main}

\end{document}